\documentclass[11pt,a4paper]{article}
\pdfoutput=1
\usepackage[hmargin=3.0cm,vmargin=3.0cm]{geometry}

\usepackage[table]{xcolor}  

\usepackage{graphicx}
\usepackage{subcaption}
\graphicspath{{figures/}} 
\usepackage{mathtools}
\usepackage{amsfonts}

\usepackage[inline]{enumitem}  
\usepackage{hyperref}
\usepackage[noabbrev]{cleveref}  

\hypersetup{
colorlinks=true,
citecolor=red
}

\usepackage{algorithm}
\usepackage[noend]{algpseudocode}
\usepackage[sort,square,numbers]{natbib}

\usepackage{hyperref}

\usepackage[group-separator={,}]{siunitx}  
\usepackage{multicol}
\usepackage{tikz}
\usepackage{smartdiagram}
\usepackage[numbers]{natbib}  
\usepackage{authblk}            

\usepackage[space]{grffile}  

\DeclareMathOperator{\E}{\mathbb{E}}
\DeclareMathOperator{\K}{\mathcal{K}}
\DeclareMathOperator{\FF}{\mathcal{F}}
\DeclareMathOperator{\LL}{\mathcal{L}}
\DeclareMathOperator{\bigO}{\mathcal{O}}
\DeclareMathOperator{\U}{\mathcal{U}}

\begin{document}\sloppy

\title{Regression-based sparse polynomial chaos for uncertainty quantification of subsurface flow models\footnote{Published in Journal of Computational Physics (2019). DOI: \href{https://doi.org/10.1016/j.jcp.2019.108909}{10.1016/j.jcp.2019.108909}}
}

\author[]{Alexander Tarakanov \footnote{Corresponding author.\\ E-mail addresses: \texttt{a.tarakanov@hw.ac.uk} (Alexander Tarakanov), \texttt{a.elsheikh@hw.ac.uk} (Ahmed H. Elsheikh).} }
\author[]{Ahmed H. Elsheikh}
\affil[]{Heriot-Watt University, United Kingdom}
\affil[]{\small School of Energy, Geoscience, Infrastructure and Society}

\maketitle

\begin{abstract}

Surrogate-modelling techniques including Polynomial Chaos Expansion (PCE) is commonly used for statistical estimation (aka. Uncertainty Quantification) of quantities of interests obtained from expensive computational models. PCE is a data-driven regression-based technique that relies on spectral polynomials as basis-functions. In this technique, the outputs of few numerical simulations are used to estimate the PCE  coefficients within a regression framework combined with regularization techniques where the regularization parameters are estimated using standard cross-validation as applied in supervised machine learning methods.

In the present work, we introduce an efficient method for estimating the PCE coefficients combining Elastic Net regularization with a data-driven feature ranking approach. Our goal is to increase the probability of identifying the most significant PCE components by assigning each of the PCE coefficients a numerical value reflecting the magnitude of the coefficient and its stability with respect to perturbations in the input data. In our evaluations, the proposed approach has shown high convergence rate for high-dimensional problems, where standard feature ranking might be challenging due to the curse of dimensionality.

The presented method is implemented within a standard machine learning library (scikit-learn~\citep{Scikit_Learn}) allowing for easy experimentation with various solvers and regularization techniques (e.g. Tikhonov, LASSO, LARS, Elastic Net) and enabling automatic cross-validation techniques using a widely used and well tested implementation. We present a set of numerical tests on standard analytical functions, a two-phase subsurface flow model and a simulation dataset for CO2 sequestration in a saline aquifer. For all test cases, the proposed approach resulted in a significant increase in PCE convergence rates.
\end{abstract}

\section{Introduction}
Uncertainty Quantification (UQ) and Uncertainty Propagation (UP) in the subsurface flow related problems have been the subject of intensive research activities over the last decades~\citep{TUGAN2018107, WANG2019105, CHEN2017328, MA20114696, UQ_CO2_1}. For instance, UQ of oil production forecasts from a given reservoir has far-reaching economical consequences~\citep{ELGSAETER20084540}. Also, the accurate risk assessment of CO2 trapping in an underground reservoir~\citep{JIA2018104} is of high importance from ecological and social perspectives~\citep{Lackner1677}.

The main challenge for UQ in subsurface flow related tasks is the complexity of the modeled physical systems~\citep{CO2_factors} and the lack of information about the rock properties that determine underground flow~\citep{Book_on_Geostatistics}. Therefore, UQ for a Quantity of Interest (QoI) is usually performed numerically through multiple evaluations of expensive reservoir simulations~\citep{DAI2019519}. This corresponds to  significant computational resources especially when dealing with a high number of uncertain parameters~\citep{UQ2} and for the cases where model high resolution is requirement~\citep{Optimization}. Several lines of research have been pursued to address this challenge. For instance, models of multiple continuum media~\citep{ZHANG2018484, LI2018127, AKKUTLU201765}, dual mesh approaches~\citep{Dual_Mesh1, Dual_Mesh2, KHOOZAN2011195}, upscaling~\citep{VASILYEVA2019, Upscaling1, Upscaling2, c98b95f9037d457990cec445e43f1fdd} and model reduction~\citep{Model_Reduction1, Model_Reduction2, Model_Reduction3, 2ae730559d3d4dfa92f1da3e3188cc79} techniques have been developed to decrease the run-time of a single simulation. Also, various surrogate modeling techniques~\citep{Proxy1, Proxy2, Proxy3} emerged in order to reduce the cost of evaluating a large number of expensive numerical simulations.

In the current manuscript, we focus on surrogate modeling approaches using PCE-based response surfaces. There are several advantages of using PCE as a proxy model. First of all, surrogate models based on sparse PCE do not require significant computational resources to compute a value at any given point within the interpolation domain as it is simply a direct polynomial function evaluation. Secondly, important statistical properties such as moments and sensitivities can be computed directly from the PCE coefficients without the need for a Monte-Carlo simulations~\citep{PCE_Stat1}. This is attributed to a special design properties of PCE that links the probability distribution of random variables with the orthogonality of polynomial basis functions~\cite{PCE_Stat1}.

Generally, two techniques could be utilized to estimate the PCE coefficients: collocation-based and regression-based methods. For collocation approaches, the QoI values are evaluated at pre-specified set of points called collocation nodes~\citep{PCE_quadrature}. These specific points are designed in such a way that the PCE coefficients can be expressed as linear combination of the QoI values, allowing for direct computation of the PCE coefficients. The optimal choice of the collocation points especially for high-dimensional problems is a subject of extensive research activities~\citep{WU2018997, XU201824, PALAR2018175}. In regression-based approaches, the PCE coefficients correspond to the solution of an error-minimization problem~\citep{PCE_Regression}. It is simple to show that the mean-square error minimization can be reduced to a linear regression problem to estimate the PCE expansion coefficients. Designing fast and accurate solution techniques to this minimization problem including various preconditioning methods is also a subject of intensive research activities~\citep{PCE_Preconditioning, ABOLGHASEMI2012999, 6484193}. ~\citet{PCE_Hybrid} developed a hybrid collocation and regression technique, where the training points for the surrogate model are generated with collocation techniques while the PCE coefficients are estimated by solving an error-minimization problem. One of the advantages of this approach is the better conditioning of the regression problem when compared to training using random samples~\citep{PCE_Hybrid}.

In generic cases, sparse collocation techniques and hybrid approaches provide accurate response surfaces using reasonable computational resources~\citep{Hosder_2010, Yan_2012, good_collocation}. However, these methods rely on evaluating the QoI at specific set of points. This strategy can be successfully adopted for UQ of oil production and CO2 storage capacity~\citep{Babaei_Paper1, Babaei_Paper2, Petvipusit2014}. However, computation of QoI values in the case of subsurface flow problems can be challenging if the collocation points correspond to extreme values of parameters that significantly affect convergence properties of the numerical scheme. Therefore, such collocation nodes can either increase computational costs of the response surface construction or reduce the overall accuracy of the surrogate model if significant numerical error is introduced to the QoI values at collocation nodes.  Additionally, for many practical problems sampling of data points can not be controlled. For instance, samples could be generated randomly (e.g. Latin hypercube sampling), or in accordance with a prescribed probability distribution~\cite{ELSHEIKH2014515}, or based on another meta-modeling technique used in combination with PCE for model stacking~\citep{Model_Selection}. Under these conditions, collocation techniques cannot be directly applied. For this reason, regression based PCE (utilized in this manuscript) have wider applicability for any set of training samples where optimal response surfaces could be built.

In regression methods, PCE coefficients are computed through the minimization of mean-square error over the training data. Therefore, for low-dimensional problems, a direct approach could be adopted. In generic setting, the number of PCE coefficients for a problem with $n$ variables can be expressed as follows:
\begin{equation}
    \label{eq:dimension_of_the_problem}
    D = D(n, d) = \binom{n+d}{n}
\end{equation}
where $D$ is the number of PCE coefficients and $d$ is the degree of polynomials used. It is simple to observe the fast growth of $D$ with both $d$ and $n$. This exponential growth of PCE coefficients imposes  significant constraints on building PCE-based response surfaces. First of all, solving the error-minimization regression problem in high dimensions is a challenging task, because of the high number of numerical operations needed till convergence. Secondly, the number of QoI values (i.e. training samples) needed for accurate estimation of the PCE coefficients increases with $D$, which corresponds to additional runs of an expensive numerical simulator. In other words, the curse of dimensionality makes it impractical to solve for PCE coefficients directly. However, for a large class of problems it was observed that PCE coefficients are sparse~\cite{HAMPTON201820, BAZARGAN2015385}. Therefore, various techniques for sparse regression can be adopted. For example, $\ell^1$ regularization techniques~\citep{Sparse_PCE_L1} can be considered as a first step towards enforcing  sparsity on the PCE regression coefficients. This approach is widely adopted and will be referred to as standard PCE~\citep{classic_sparse_PCE} in the rest of this manuscript. Further dimension reduction could be achieved through fitting both the data and the QoI derivatives at the training points~\citep{PCE_Derivatives2}. The additional information from the gradients increases the quality of PCE response surface~\citep{PCE_Derivatives1}. Unfortunately, for many problems it is not possible to obtain the gradient information at the training points. Another line of research focuses on reducing the problem dimension by using advanced methods for solving nonlinear regression problems. For instance, sparse PCE coefficients can be computed efficiently through the application of support vector regression~\citep{PCE_SVR} or preconditioned conjugate gradient~\citep{PCE_PCG} techniques. Another direction of development relies on coupling the iterative solvers with algorithms for ranking the importance of the basis polynomials (e.g. orthogonal matching pursuit~\citep{PCE_Regression}) or ranking based on the impact on the residual~\citep{PCE_Residual_Based_Ranking}. Further reduction of dimension could also be achieved by adaptive truncation of the spectrum of the expansion. For instance, it has been observed empirically for a broad class of problems, that higher order interactions between the polynomial basis from different dimension have less impact on the quality of the response surface when compared to the one dimensional low order polynomials. This empirical observation is the foundation for hyperbolic truncation techniques~\citep{Hyperbolic_Truncation_Scheme}. Moreover, the performance of all regression-based approaches could be improved by transformation of the input variables (e.g. scaling, normalization). For example, variable rotations~\citep{PCE_Rotations} or generic linear transformations~\citep{PCE_Linear_Combination_of_Variables} could significantly reduce the complexity of the error minimization problem corresponding to finding the PCE coefficients.

In the current paper, we focus on further improvement of dimension reduction techniques for regression based PCE. We present a novel iterative approach for solving the error minimization problem. We introduce a new data-driven ranking procedure for sequential identification of the most significant PCE basis functions with the closest relation to the interpolated QoI values. The ranking procedure is based on the correlation between the basis functions and the QoI values penalized by factors that measure the sensitivity of the corresponding coefficient to the noise in the input data. The aim of the introduction of correction/penalty factors is to avoid overestimation of the significance of a given polynomial basis function due to occasional location of data-points. The introduced ranking approach enables us to determine the most significant PCE terms and subsequently solving a reduced regression problem at each iteration. The new method could be easily combined with various regularization techniques.

The proposed approach has been integrated in scikit-learn~\citep{Scikit_Learn}, a widely used machine learning library. This integration enables uniform testing of a huge variety of techniques such as Lasso, Lars and Elastic Net~\citep{Regularized_Linear_Regression} in order to formulate and solve the regularized regression problem. We implement PCE as an input feature transformation using machine learning terminology. Therefore, PCE can be naturally included in any machine-learning pipeline allowing one to combine different methods for variable transformation with advanced cross-validation techniques. Moreover, this implementation allows for an easy comparisons to alternative machine-learning techniques (e.g. Random Forests, Support Vector Machines). In the numerical evaluation section, we compare the proposed approach to classical methods for sparse PCE namely, the Orthogonal Matching Pursuit (OMP) and Least Angular Regression (LARS). We consider four data-sets for evaluation. The first two data-sets are generated using analytical functions and the last two data-sets are based on subsurface simulations of fluid flow in porous media. In all the test cases, extensive comparisons are performed in terms of Mean-Square Error (MSE) using a hold-out (aka. validation) set of points following the best practices in the machine learning literature.

The rest of this manuscript is organized as follows: In the following section, a general introduction to PC is presented followed by the proposed ranking procedure. In section~\ref{sec:numerical_examples} we present a set of numerical examples. Finally, the conclusion of our work is presented in section~\ref{sec:conclusions}.

\section{Methodology} \label{sec:Methodology}
Polynomial chaos expansion PCE is a meta-modeling technique that relies on orthogonal polynomials. One of the main advantages of PCE when compared to alternative surrogate modeling techniques is the ability to estimate the QoI sensitivity to given combination of variables through simple analytical formulae. This is only possible due to the close relation between the orthogonality of basis polynomials and the probability distribution of the input variables. 
This relation is explained in subsection~\ref{subsec:overview_of_polynomial_chaos}, along with an overview of basic ideas of PCE. The proposed reordering of PCE basis is then introduced in subsection~\ref{subsec:regression_based_pce}.

\subsection{Basics of Polynomial Chaos}
\label{subsec:overview_of_polynomial_chaos}
The essence of PCE is the relation between the statistics of input data and orthogonality of the utilized basis polynomials. The relation concerned gives a powerful tool for calculating the PCE coefficients and for further statistical analysis of the data. We first explain this relation for the single-variate case and then extend this formulation to multi-variate cases. Additionally, examples of applying this concept to study the statistical properties of PCE are presented.

For single-variate function $f(x)$, a PCE is defined as series of orthogonal polynomials:
\begin{equation}
    \label{eq:generic_1d_PC_expansion}
    f(x) = \sum_{\alpha} \textit{c}_{\alpha}p_{\alpha}(x)
\end{equation}
where $p_{\alpha}(x)$ is an orthogonal single-variate polynomial with the index $\alpha$ and $c_{\alpha}$ is the corresponding PCE coefficient. The specific type of utilized orthogonal polynomials is not of a principal importance in the definition introduced in Eq.~\eqref{eq:generic_1d_PC_expansion}. Therefore, PCE can be naturally formulated for all well-known families of orthogonal polynomials. For example Hermite, Legendre and Chebyshev polynomials~\citep{GUO2019129}.

The analysis of the PCE relies on the orthogonality of the basis polynomials, which is introduced through the notion of an inner product defined as following:
\begin{equation}
    \label{eq:generic_inner_product}
    \langle g_1, g_2 \rangle = \int_{-\infty}^{+\infty} \K(x)g_1(x)g_2(x) \mathrm{d}x
\end{equation}
where $g_1(x)$ and $g_2(x)$ are certain square-integrable functions and $\K(x)$ is a non-negative function referred to as the kernel function or simply the kernel. Classical families of orthogonal polynomials are related to a specific form of the kernel function. For instance, Hermite polynomials correspond to a kernel function identical to Gaussian distribution function with zero mean and unit variance~\citep{PC_Hermite}:
\begin{equation}
    \label{eq:Hermite_kernel}
    \K(x) = \frac{1}{\sqrt{2\pi}} e^{- {{x^2}/{2}}}
\end{equation}
For a generic case, the PCE basis functions are constructed by applying Gram-Schmidt orthogonalization to the set of monomial functions (e.g. $1,~x,~x^2,~\dots$)~\citep{Book_on_Orthogonal_Polynomials}. Therefore, PCE techniques could be naturally extended to any arbitrary kernel functions $\K(x)$.

A central idea of PCE is the statistical interpretation of $\K(x)$ as probability density function for a given random variable~\citep{CORTES20171}. This interpretation allows one to reformulate the inner product defined in Eq.~\eqref{eq:generic_inner_product} in terms of expectations:
\begin{equation}
    \label{eq:generic_inner_product_and_probability}
    \langle g_1, g_2 \rangle = \int_{-\infty}^{+\infty} \K(x)g_1(x) g_2(x) \mathrm{d}x = \E[g_1, g_2]
\end{equation}
In the setting, the orthogonality of polynomials $p_{\alpha}(x)$ with respect to the inner product can be reformulated as:
\begin{equation}
    \label{eq:orthogonality_def}
    \langle p_{\alpha}, p_{\beta} \rangle = \E[p_{\alpha}, p_{\beta}] = \|p_{\alpha}\|^2 \delta_{\alpha\beta}
\end{equation}
where $\delta_{\alpha\beta}$ is a Kronecker symbol. In the present work, we consider orthonormal polynomials with $\|p_{\alpha}\|^2 = 1$ in order to simplify the numerical analysis of PCE. Therefore, Eq.\eqref{eq:orthonormality_def} can be transformed as follows:
\begin{equation}
    \label{eq:orthonormality_def}
    \langle p_{\alpha}, p_{\beta} \rangle = \E[p_{\alpha}, p_{\beta}] = \delta_{\alpha\beta}
\end{equation}
Moreover, the basis polynomials orthogonality can be used to estimate the PCE coefficients:
\begin{equation}
    \label{eq:PCE_coef_generic}
    c_{\alpha} = \langle f, p_{\alpha} \rangle = \E[f, p_{\alpha}]
\end{equation}

For multi-variate functions, similar analysis could be performed through the introduction of the tensor-product concept where the set of multivariate basis functions is formed as products of single-variate polynomials:
\begin{equation}
    \label{eq:tensor_product}
    p_A(\textbf{x}) = p_{\alpha_1}^{(1)}(x_1) p_{\alpha_2}^{(2)}(x_2) \dots p_{\alpha_n}^{(n)}(x_n)
\end{equation}
where $\alpha_k$ is the degree of single-variate polynomial, $p_{\alpha_k}^{(k)}(x_k)$ is a uni-variate polynomial that depends only on the $k$-th coordinate of the vector $\mathbf{x}$. The degree of polynomial $p_A(\textbf{x})$ is defined as:
\begin{equation}
    \label{eq:degree_of_polynomials}
    \text{deg}(p_A(\textbf{x})) = \sum_k \text{deg}(p_{\alpha_k}^{(k)}(x_k)) = \sum_k \alpha_k
\end{equation}
Similar to Eq.~\eqref{eq:generic_1d_PC_expansion}, the PCE of multivariate function $f(\textbf{x})$ is defined as:
\begin{equation}
    \label{eq:generic_PC_md_expansion}
    f(\textbf{x}) = f(x_1,\dots,x_n) = \sum_A c_A p_A(\textbf{x})
\end{equation}
where $c_A$ is the PCE coefficient corresponding to polynomial basis function with multi-index $A$.
The inner product in multi-dimensional case is defined as:
\begin{equation}
    \label{eq:md_inner_product}
    \langle g_1, g_2 \rangle = \int \K(\mathbf{x}) g_1(\mathbf{x}) g_2(\mathbf{x}) \mathrm{d}\mathbf{x}
\end{equation}
where $\K(\mathbf{x})$ is a multi-variate kernel function and $g_1(\mathbf{x})$, $g_2(\mathbf{x})$ are certain square-integrable functions. It is important to note that the polynomial basis functions obtained by tensor multiplications inherit the orthogonality and orthonormality from single-variate PCE if the multi-variate kernel function $\K(\mathbf{•}{x})$ equals the product of single-variate kernel functions:
\begin{equation}
    \label{eq:kernel_product}
    \K(\textbf{x}) = \K_1(x_1) \cdots \K_n(x_n)
\end{equation}
where $\K_1(x_1), \cdots , \K_n(x_n)$ are single-variate kernel functions. From a probabilistic point of view, this is equivalent to the mutual independence of the coordinates of the vector $\textbf{x}$.

The inner product defined in Eq.~\eqref{eq:md_inner_product} can be utilized to derive an expression for the PCE coefficients similar to Eq.~\eqref{eq:PCE_coef_generic}:
\begin{equation}
    \label{eq:PCE_coef_generic_multi_variate}
    c_{A} = \langle f, p_{A} \rangle = \E[p_A, f]
\end{equation}
The relation between the input data statistics and the polynomial basis orthogonality can be used to derive analytical expressions for the mean, variance and Sobol' indices of the function $f(\textbf{x})$. For example, the mean can be estimated by:
\begin{equation}
    \label{eq:mean_calculation}
    \E[f] = \langle 1, f \rangle = \sum_A c_A \langle 1, p_A \rangle = c_{0, \dots, 0} = c_0
\end{equation}
Where $c_{0, \dots, 0}$ is the constant polynomial coefficient. In the present work, we simplify the notations and use $c_{0}$ instead of $c_{0, \dots, 0}$. The mean-square deviation can be calculated in the similar fashion:
\begin{equation}
    \label{eq:variance_calculation}
    \sigma^2 = \E[(f-c_0), (f-c_0)] = \sum_{A_1, A_2 \neq 0} c_{A_1} c_{A_2} \delta_{A_1 A_2} = \sum_{A>0} c_A^2
\end{equation}
Calculation of other quantities for sensitivity analysis and UQ such as partial variances and Sobol' indices could be performed naturally with PCE. A partial standard deviation represents the sensitivity to a given combination of variables. It is defined as the standard deviation of the function $f(\mathbf{x})$ averaged with respect to certain collection of variables~\citep{PALAR2018175}:
\begin{equation}
    \label{eq:partial_variance}
    \sigma^2_{r_1, \dots, r_k}(f) = \sigma^2(\E_{t_1, \dots, t_{n-k}}[f])
\end{equation}
where $\sigma_{r_1, \dots, r_k}$ is the standard deviations with respect to the components of the vector $\mathbf{x}$ with indices $r_1, \cdots, r_k$ and $\E_{t_1, \dots,t_{n-k}}[f]$ is the average with respect to components of the vector $\mathbf{x}$ with indices $t_1, \cdots, t_{n-k}$ that form a complement to $r_1, \cdots, r_k$~\citep{PCE_Stat1}, $n$ is the dimension of $\mathbf{x}$ and $k$ is a certain integer number from $1$ to $n$. Sobol' indices are commonly used as a measure for sensitivity and are defined as the normalized partial standard deviations:
\begin{equation}
    \label{eq:sobol_index}
    S_{r_1, \dots, r_k}(f) = \frac{ \sigma^2_{r_1, \dots, r_k}(f)} {\sigma^2(f)}
\end{equation}
For the response function $f$ with a PCE representation, the partial standard deviations can be calculated in a similar fashion as the normal standard deviation defined in Eq.~\eqref{eq:variance_calculation} following~\citep{PALAR2018175}:
\begin{equation}
    \label{eq:partial_variance_pc}
    \sigma^2_{r_1, \dots, r_k}(f) = \sum_{\alpha_{r_l} > 0, \alpha_{t_j} = 0 } c_A^2
\end{equation}

The relation between orthogonality of polynomial basis functions and the probability distribution of input data has an important consequence on the numerical calculation of the PCE coefficients. In practice, for regression based response surfaces, the PCE coefficients for a given function $f(x)$ can be computed for given input data through the minimization of the mean-square error (MSE) functional:
\begin{equation}
    \label{eq:mean_square_functional}
    \FF(\mathbf{c}) = \sum_i \frac{(y_i - \sum_A \textit{c}_Ap_A(\textbf{x}_i))^2} {N}
\end{equation}
where $\textbf{x}_i$ is the $i^{\text{th}}$ vector of input variables, $N$ is the number of data points and $y_i$ is the value of the function $f$ at the point $\textbf{x}_i$. In the present work the spectrum of PCE is truncated to a certain polynomial degree $d$. Therefore, the dimension of $\mathbf{c}$ is given by Eq.~\eqref{eq:dimension_of_the_problem}.

It is simple to see that minimizing the functional defined in Eq.~\eqref{eq:mean_square_functional} is equivalent to solving a system of linear equations:
\begin{equation}
    \label{eq:linear_problem}
    \textbf{M}_{AB}c_B = \textbf{V}_A
\end{equation}
where the square matrix $\mathbf{M}$ and vector $\mathbf{V}$ are defined as:
\begin{equation}
    \label{eq:matrix_and_vector_notations}
        \textbf{M}_{AB} = \sum_i \frac{p_A(\textbf{x}_i) p_B(\textbf{x}_i)}{N}, \qquad
        \textbf{V}_A = \sum_i \frac{y_i p_A(\textbf{x}_i)}{N}
\end{equation}
The relation between the basis orthogonality and the statistical distribution of the input data imposes several constraints on the value of the matrix $\textbf{M}$ and the vector $\textbf{V}$. If the data is sampled in agreement with the probability distribution determined by the kernel function defined in Eq.~\eqref{eq:kernel_product}, then the matrix $\textbf{M}$ should converge to a unit matrix:
\begin{equation}
    \label{eq:estimate_for_the_matrix_element}
    \textbf{M}_{AB} = \sum_i \frac{p_A(\textbf{x}_i) p_B(\textbf{x}_i)}{N} = \E[p_Ap_B] + \bigO \bigg ( \frac{1}{\sqrt{N}}\bigg ) = \delta_{AB} + \bigO \bigg ( \frac{1}{\sqrt{N}}\bigg )
\end{equation}
where the term $\bigO \big( \frac{1}{\sqrt{N}}\big)$ represents the convergence in accordance with the law of large numbers~\citep{Probability_Textbook}. Similar reasoning could be applied to the vector $\textbf{V}$ showing the close correlation between the data and the basis functions:
\begin{equation}
    \label{eq:estimate_for_the_vector}
    \textbf{V}_A = \sum_i \frac{p_A(\textbf{x}_i) y_i}{N} = \E[y(x)p_A(x)] + \bigO \bigg ( \frac{1}{\sqrt{N}}\bigg ) = \textit{c}_A + \bigO \bigg ( \frac{1}{\sqrt{N}}\bigg )
\end{equation}
Eq.~\eqref{eq:estimate_for_the_matrix_element} and Eq.~\eqref{eq:estimate_for_the_vector}, simply means that the coefficients $\textbf{c}$ minimizing the MSE functional defined in Eq.~\eqref{eq:mean_square_functional} is close to the correlation vector $\textbf{V}$ if a sufficient number $N$ of training data points is available. Moreover, the difference between $\textbf{V}$ and $\textbf{c}$ can be estimated as follows:
\begin{equation}
    \label{eq:inequality_for_coefficients}
    |V_A - c_A| \leq \frac{k_A}{\sqrt{N}}
\end{equation}
Where $k_A$ is a positive number. In other words, $\textbf{V}$ provides a reasonable approximation for $\textbf{c}$ if a sufficient number of data-points is available. We utilize this observation to introduce a novel ranking-based approach to approximate the PCE coefficients as described in the next subsection.

\subsection{Ranking based sparse PCE}
\label{subsec:regression_based_pce}
In the present work, we estimate the PCE coefficients by minimizing the mean-square error functional defined in Eq.~\eqref{eq:mean_square_functional}. It is well-known that a straight-forward minimization of mean square errors could provide an unstable solution or a response surface that is not quite accurate at points that are not included in the training data-set. Therefore, we utilize a mixed $\ell_1$ and $\ell_2$ regularization technique known as Elastic Net model~\citep{Elastic_Net1} (i.e., combined Tikhonov and Lasso regularization). This results in a regularized functional for error minimization of the following form:
\begin{equation}
    \label{eq:minimization_problem}
    \textbf{c} = \underset{\textbf{c}}{\arg\min} \LL(\textbf{c}) = \underset{\textbf{c}}{\arg\min} \bigg( \FF(\textbf{c}) + \lambda_1 \sum_{A} |c_A| + \lambda_2 \sum_{A} c_A^2\bigg)
\end{equation}
where $\LL(\textbf{c})$ is a functional for minimization and $\lambda_1, \lambda_2$ are hyperparameters that could be tuned in order to maximize the quality of the surrogate model. In the present work, $\lambda_1$ and $\lambda_2$ are determined through cross-validation.

We utilize a coordinate descent algorithm~\citep{Regularized_Linear_Regression} in order to find the solution for the minimization problem defined in Eq.~\eqref{eq:minimization_problem}. This is an iterative algorithm that sequentially updates the solution vector $\textbf{c}$ by minimizing the functional $\LL(\textbf{c})$ with respect to one of the coordinates at each step as summarized in Algorithm~\ref{alg:coordinate_descent}.

\begin{algorithm}[!h]
\caption{Coordinate descent}\label{alg:coordinate_descent}
\begin{algorithmic}
\State $\textbf{c} = 0 $ \Comment{Set vector of parameters to zero}
\While{$\Delta \LL > \varepsilon $}\Comment{Iterate while change in $\LL(c)$ is significant}
\State Select a value $k$ from $1$ to $\text{dim}(\textbf{c})$ \Comment{Select one of the coordinates}
\State $c_{k} = \underset{c_k}{\arg\min} \LL(\mathbf{c})$ \Comment{Minimize with respect to single parameter}
\State Update $\Delta \LL$
\EndWhile
\State \textbf{return} $\mathbf{c}$
\end{algorithmic}
\end{algorithm}

One of the essential parts in Algorithm~\ref{alg:coordinate_descent} is the selection of the next component for update. Classical approaches include: random selection or selection based on the cyclic order on the set of components~\citep{Regularized_Linear_Regression}. In the present work, we introduce a novel scheme for reordering the polynomial basis functions that increases the algorithm convergence rate and increases the response surface quality when utilizing small number of training samples. The aim of the reordering procedure is to identify the polynomial basis functions with the highest PCE coefficients in order to determine its values first. It should be noted that the assumption about the agreement between sampling of training data and orthogonality of basis polynomial functions Eq.~\eqref{eq:generic_inner_product_and_probability} is of principal importance for the proposed reordering procedure. For the cases where this assumption is violated, data transformation techniques should be applied before using the proposed reordering approach. For instance, the desired distribution of input variables can be achieved through quantile transformation~\citep{Quantile} or Rosenblatt transformation~\citep{rosenblatt1952}.

The reordering technique utilizes a ranking of PCE coefficients inspired by Eq.~\eqref{eq:estimate_for_the_vector}, which states that the vector of moments is close to the actual PCE coefficients given a sufficient number of training points. However, for certain polynomial basis functions the difference between $c_A$ and $V_A$ can be significant leading to an overestimation of the importance of those components due to the lack of the available data, which can be considered as a noise. In order to address this issue, we introduce a ranking of polynomial basis functions in a form of the signal-to-noise ratio which is a correlation coefficient divided by a correction factor that quantifies the sensitivity of a given PCE coefficient to the data noise.

Two sources of noise are considered in the current work: noise in the values of QoI and noise in the deviation of matrix $\mathbf{M}$ due to the random sampling of the training data. In order to quantify both sources of noise, we perform two series of Monte-Carlo simulations. In the first series of Monte-Carlo simulations, the sensitivity of the correlation vector $\mathbf{V}$ to the QoI values is estimated. For that purpose, we introduce random perturbations $\theta_i$ to each of the training data-points. In the present study, the noise part is sampled from a normal distribution with zero mean and unit variance. The correlation of the basis polynomial $p_A(\mathbf{x})$ with perturbed data to the QoI is estimated using:
\begin{equation}
    \label{eq:U_definition}
    {\bf U}_A = \sum_i \frac{(y_i+\theta_i) p_A(\textbf{x}_i)}{N}
\end{equation}
The mean-square deviation $\sigma_{Y,A}$ of ${\bf U}_A$ from $\bf{V}_A$ is considered as a measure for stability:
\begin{equation}
    \label{eq::sensitivity_measure_for_Y}
    \sigma_{Y,A} = \sqrt{\E_{\theta}[({\bf U}_A-{\bf V}_A)^2]}
\end{equation}
where the mean $\E_{\theta}$ is taken over several realization of $\theta$.

The second series of Monte-Carlo simulations is performed to quantify the stability with respect to the location of training points. As long as the location of training data points $\tilde{\textbf{x}}$ is considered as a random parameter, a set of $N$ points is generated at each Monte-Carlo simulation. Then the mean-square deviation $\sigma_{X, A}$ of ${\mathbf M}_{AA}$ from the unit matrix can be computed numerically as follows:
\begin{equation}
    \label{eq:sensitivity_measure_for_X}
    \sigma_{X,A} = \sqrt{E_{\tilde{\textbf{x}}}[({\mathbf M}_{AA} - \mathbf{I})^2]}
\end{equation}
where the mean $\E_{\tilde{\textbf{x}}}$ is taken over a number of realizations of $\tilde{\mathbf{x}}$.
Finally, the ranking coefficient for the basis polynomial $p_A(\mathbf{x})$ is defined as:
\begin{equation}
    \label{eq:ranking_coefficient_final}
    r_A = \frac{1} {\sqrt{\sigma_{Y,A}^2 + \sigma_{X,A}^2}} \frac{|\textbf{V}_A|} {H}
\end{equation}
where parameter $H$ is introduced for normalization purposes. The value of $H$ is given by the expression:
\begin{equation}
    \label{eq:D_coef}
    H = \frac{1}{2}\min_A|\textbf{V}_A| + \frac{1}{2}\max_A|\textbf{V}_A|
\end{equation}
where minimum and maximum are taken over all values of multi-index $A$. In this work we consider high values of $r_A$ as an indicator of a high value of the corresponding PCE coefficient.

In the present work, the ranking parameter $r_A$ is used within the coordinate descent Algorithm~\ref{alg:coordinate_descent} to select the next PCE coefficient for updates. Therefore, we solve iteratively for PCE coefficients by performing the following steps sequentially: ranking of basis functions based on the residual $\eta^{(k)}$ at the step $k$, select the first $N_B$ basis functions and solve for the corresponding PCE coefficients using coordinate descent method. These steps are combined in Algorithm~\ref{alg:PCE_solver}. In our numerical testing, we set $N_B=5$ based on some initial testing. However, a more rigorous approach could utilize cross validation to select the optimal number of $N_B$.

\begin{algorithm}[h]
\caption{Ranking based sparse PCE solver}\label{alg:PCE_solver}
\begin{algorithmic}[1]
\State k=0 \Comment{Set iteration counter to zero}
\State $\eta^{(0)}$ = y \Comment{Residual equals to initial data}
\State $\textbf{c} = 0 $ \Comment{Set vector of parameters to zero}
\While{$\Delta \LL > \varepsilon $} \Comment{Iterate while change in $\LL(c)$ is significant}
\State $\textbf{r} = \textbf{r}(\eta^{(k)})$ \Comment{Compute ranking}
\State Select $A_1, \cdots , A_{N_B}$ \Comment{indices of first $N_B$ components with highest rank}
\State Solve $\Delta \mathbf{c} = \underset{\Delta \mathbf{c}}\arg\min(\LL(\Delta c_{A_1}, \cdots , \Delta c_{A_{N_B}}))$ with respect to selected components \Comment{Use Algorithm~\ref{alg:coordinate_descent}}
\State Update coefficients: $\mathbf{c} = \mathbf{c} + \mathbf{\Delta c}$
\State Update residual: $\eta^{(k+1)} = y - \sum_A c_A p_A(\mathbf{x})$
\State $k = k+1$
\EndWhile
\State \textbf{return} $c$
\end{algorithmic}
\end{algorithm}

\section{Numerical Examples}
\label{sec:numerical_examples}
In this section, the proposed ranking based sparse PCE is evaluated on four test cases. The first test case is the Ishigami function~\citep{Modified_Ishigami}, the second test case is a ten-dimensional Ackley function, the third case is a waterflooding problem with uncertain permeability field and the forth test case utilizes a data-set from simulations of CO2 injection~\citep{UQ_CO2_2}. In all test cases, the proposed PCE approach is compared to two standard techniques for sparse regression-based PCE: Least Angular Regression~\citep{efron2004} and the Orthogonal Matching Pursuit (OMP) algorithm~\citep{singaravelu:06:eurosys}. The numerical implementations are all based on scikit-learn, a machine library including with standard implementation of the LARS, OMP and coordinate descent algorithm. Moreover, cross-validation tools within this library are used to select the optimal regularization parameters for the Elastic Net functional defined in Eq.~\eqref{eq:minimization_problem}.

\subsection*{Test case 1: Ishigami Function}
Ishigami function is one of the standard benchmarks~\citep{Modified_Ishigami}:
\begin{equation}
    \label{eq:modified_Ishigami_Function}
        y = 1 + \frac{1 + \pi^4/10 + \sin(\pi x_1) + 7\sin^2(\pi x_2) + 0.1 (\pi x_3)^4 \sin(\pi x_1)}{9 + \pi^4/5}
\end{equation}
This three dimensional function shows a strong nonlinear behavior and is commonly used as a test function for evaluating different response surface techniques. Typically, the evaluation domain is $[-\pi, \pi]^3$. In the present work, PCE with Legendre polynomials is used because of the finite length of the interval concerned. We have rescaled the input parameters linearly to the interval $[-1,1]$, given that the Legendre polynomials are defined over the interval $[-1,1]$. For each of the rescaled variables, a uniform distribution over the interval $[-1,1]$ is assumed. We consider two training sets of $200$ samples and $2000$ samples. Another set of $2000$ points uniformly sampled over the cube $[-1,1]^3$ is utilized for out-of-sample MSE calculations (aka. test set). We construct a PCE of polynomial functions up to degree $d=10$. The aim of the example is to compare the convergence rates of the proposed ranking based PCE approach against the standard sparse regularization techniques (i.e. LARS and OMP), while increasing the number of free coefficients $N_D$ available for fitting by these iterative techniques. In the case of LARS and OMP, the value of $N_D$ is well-defined. For the proposed ranking based approach, $N_D$ is defined as:
\begin{equation}
    \label{eq:number_of_degrees_of_freedom}
    N_D = N_I N_B
\end{equation}
where $N_I$ is the number of iterations and $N_B$ is the number of PCE coefficients that can be modified by coordinate descent solver after each ranking update in Algorithm~\ref{alg:PCE_solver}. It should be emphasized that PCE coefficients are selected solely based on ranking. In other words, the overlapping with previously selected PCE coefficients can occur. Therefore, the value of $N_D$ given by Eq.~\eqref{eq:number_of_degrees_of_freedom} is a conservative upper bound for the total number of polynomial basis functions involved in the response surface construction (i.e. with non-zero coefficient).

\begin{figure}[H]
\centering
    \begin{subfigure}[b]{0.45\textwidth}
        \centering
        \includegraphics[width=1.0\linewidth]{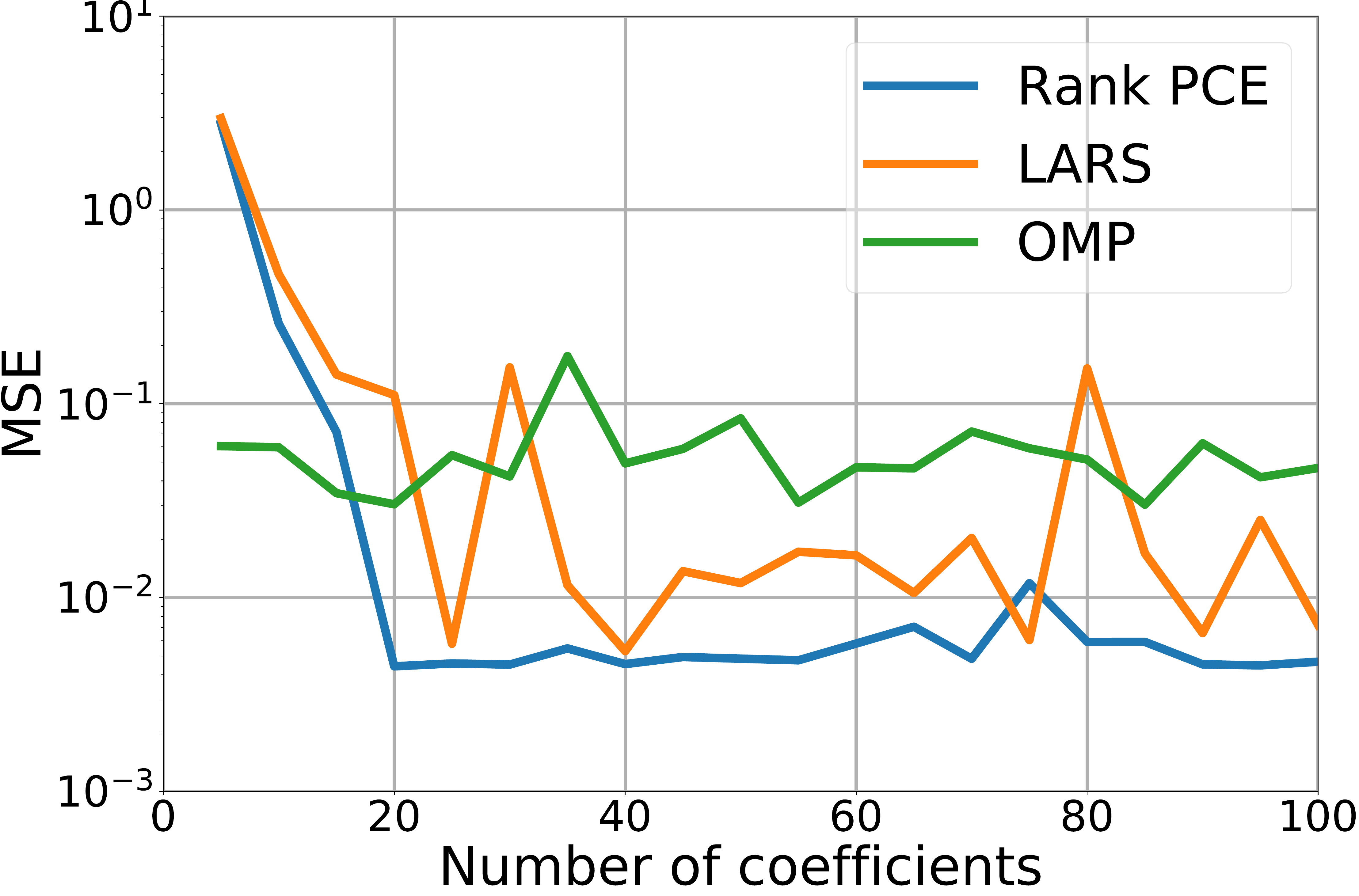}
        \caption{$200$ training points}
        \label{fig:Ish_Convergence_n_200}
    \end{subfigure}
    \begin{subfigure}[b]{0.45\textwidth}
        \centering
        \includegraphics[width=1.0\linewidth]{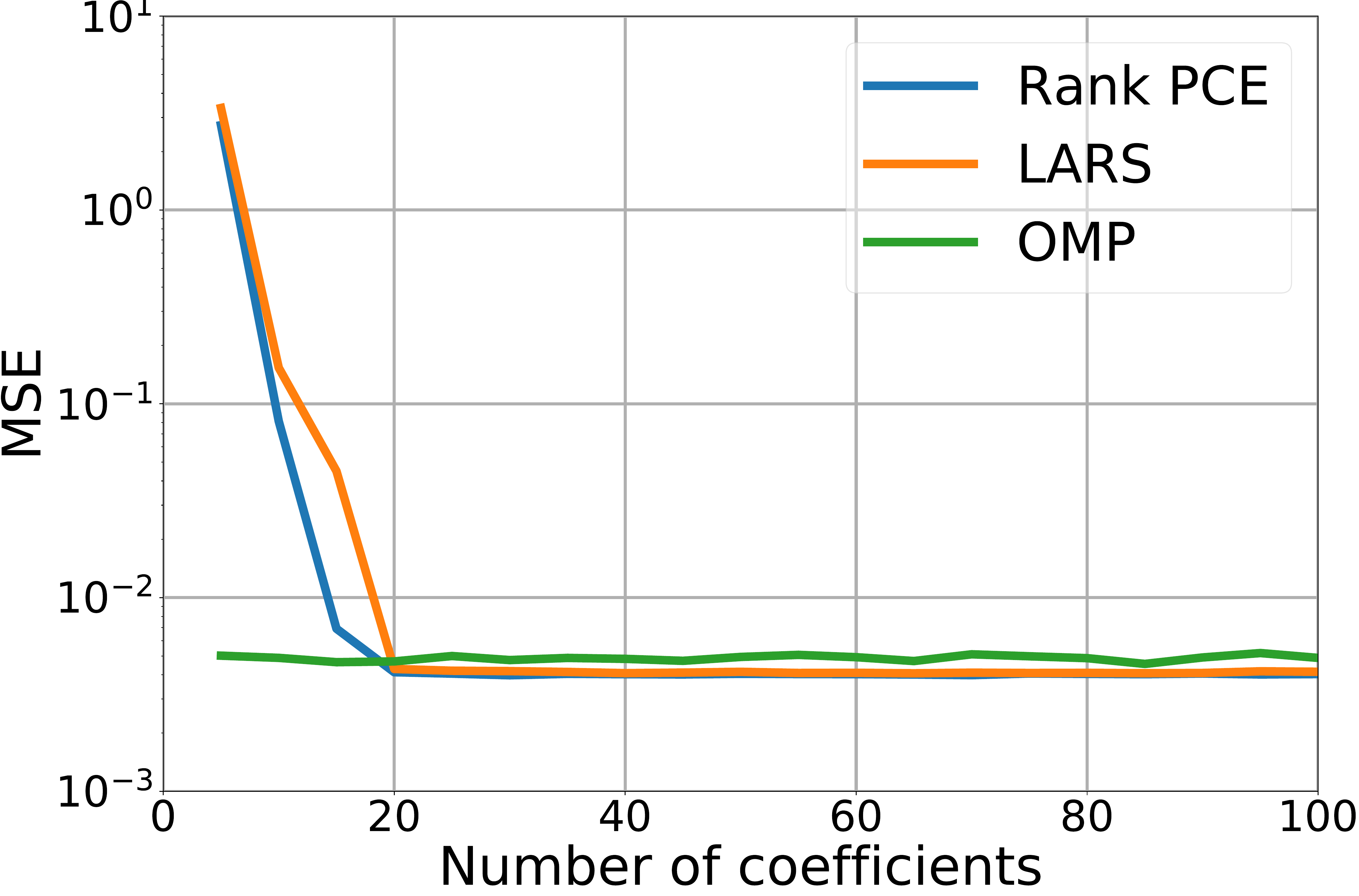}
        \caption{$2000$ training points}
    \label{fig:Ish_Convergence_n_2000}
    \end{subfigure}
    \caption{Mean-square error on the test data set versus the number of free coefficients for the Ishigami function.}
    \label{fig:Ish_Convergence}
\end{figure}

The numerical level of tolerance has been set to $10^{-6}$ in all numerical schemes. Fig.~\ref{fig:Ish_Convergence_n_200} and Fig.~\ref{fig:Ish_Convergence_n_2000} shows the MSE for each method versus the number of free coefficients $N_D$ for $200$ and $2000$ training points, respectively.

The results presented in Fig.~\ref{fig:Ish_Convergence}, shows that response surface built using the ranking based sparse PCE is of higher quality when compared to those obtained by the standard LARS or OMP algorithm, especially when the size of training data is limited. However, all three techniques perform similarly in the case with higher number of training data-points as shown in Fig.~\ref{fig:Ish_Convergence_n_2000}. This is a major advantage when collecting training samples corresponds to running computationally expensive simulations.

\subsection*{Test case 2: Ackley Function}
In this example, we build a response surface for a $10$-dimensional Ackley function~\citep{Ackley_Function} of the form:
\begin{equation}
    \label{eq:Ackley_function}
        y = -20 \exp\Bigg ( -0.2 \bigg ( {\frac{1}{n} \sum_{k=1}^n x_k^2} \bigg )^{1/2} \Bigg ) - \text{exp}\Bigg ( {\frac{1}{n} \sum_{k=1}^n \cos{2 \pi x_k}} \Bigg ) + 20 + \exp \big( 1 \big)
\end{equation}
where $n$ is the dimension of the input vector. This function shows a strong nonlinear behavior with plenty of local minimums and is frequently used as a benchmark for optimization algorithm. The setup of the current numerical example is similar to the first test case.  However, we assume that each of the input variables is uniformly distributed in the interval $[-5, 5]$. Legendre polynomials are utilized as basis function for the PCE. Therefore, input rescaling is applied to map all input variables to the interval $[-1, 1]$. In other words, we consider data to be uniformly distributed in the cube $[-1,1]^{10}$. Similar to the first test case, two training sets sizes are considered ($200, 2000$ samples) and $2000$ samples points uniformly distributed in the cube $[-1,1]^{10}$ are set aside as a test set for calculating the out-of-sample MSE. We truncate the PCE spectrum to polynomial functions up to degree $d=8$.

\begin{figure}[H]
\centering
    \begin{subfigure}[b]{0.45\textwidth}
        \centering
        \includegraphics[width=1.0\linewidth]{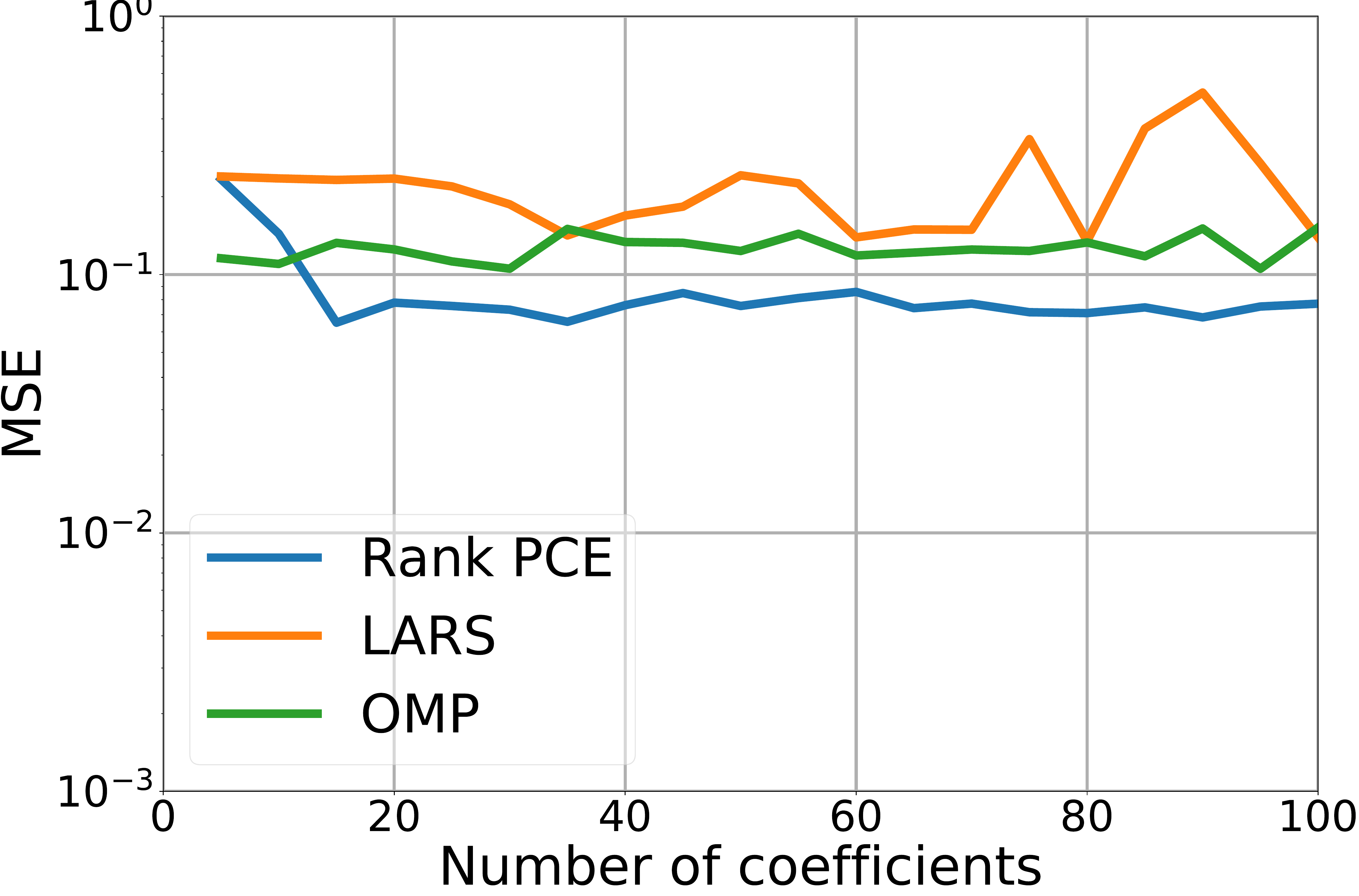}
        \caption{$200$ training points}
        \label{fig:Ackley_Convergence_n_200}
    \end{subfigure}
    \begin{subfigure}[b]{0.45\textwidth}
        \centering
        \includegraphics[width=1.0\linewidth]{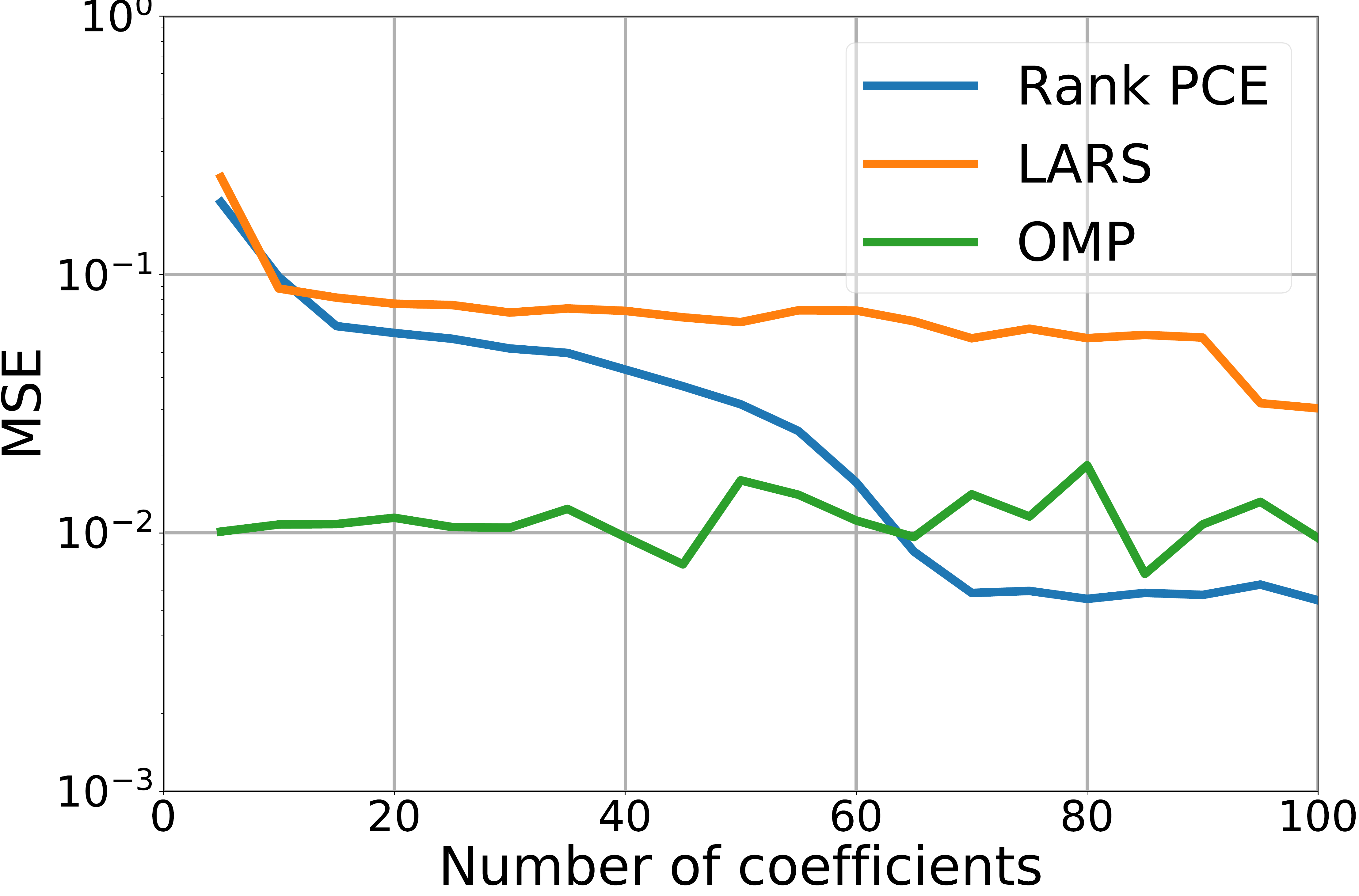}
        \caption{$2000$ training points}
    \label{fig:Ackley_Convergence_n_2000}
    \end{subfigure}
    \caption{Mean-square error on the test data set versus the number of free coefficients for a ten dimensional Ackley function.}
    \label{fig:Ackley_Convergence}
\end{figure}

Figure~\ref{fig:Ackley_Convergence} shows the MSE convergence for the ranking based sparse PCE versus LARS and OMP algorithms. Similar to the first test case, the proposed approach produces a response surface of higher quality than LARS or OMP if the size of training data is limited as shown in Fig.~\ref{fig:Ackley_Convergence_n_200}, while all techniques perform similarly a higher number of training data-points is used and a high number of polynomial basis functions is utilized as shown in Fig.~\ref{fig:Ackley_Convergence_n_2000}.

\subsection*{Test case 3: Waterflooding problem}
In the present test case, we evaluate the developed PCE approach on an uncertainty propagation for a waterflooding problem with uncertain permeability field. Dimension reduction using PCA technique is applied to the spatial field as an effective parametrization techniques~\citep{MA20117311}. Waterflooding is a commonly used process within the petroleum industry for achieving higher hydrocarbon recovery rates. The essence of this approach is to inject water through a number of wells in a given reservoir in order to displace the existing oil and increase the oil production from another set of wells, which are commonly spatially scattered to surround the injection wells. The increased productivity is observed until the injected water starts to appear at the production wells. Thus estimating when water will appear at the production wells (commonly known as the water breakthrough time~\citep{AHMED2019901}) is of significant practical importance. We note, that this time is commonly measured in terms of volume of water injected relative to the total reservoir pore volume (PVI). Prediction of the water breakthrough time $t_b$ is of high importance for hydrocarbon field development because of the economical effects associated with it. In addition, $t_b$ is highly sensitive to the spatial distribution of reservoir properties~\citep{HENDERSON2017178} (e.g. porosity, permeability) which are highly uncertain because of lack of observations. Moreover, reliable forecast for hydrocarbon production rate after the water breakthrough is significant for economical decisions. Therefore, in the present test case we develop a surrogate model for the production rate at late stages of the well life. In particular, we focus on the oil production rate $q_{\text{oil}}$ when $60\%$ of PVI has been injected.

The waterflooding system is modeled via mass and momentum conservation laws coupled with Darcy's law. A simplified model for waterflooding is utilized where flow of two incompressible fluids (water and oil) is considered. In this setting, we are interested in predicting the spatial distribution of volumetric fractions $s_a$ (saturation) of each of the fluids. The index $a$ could be replaced by either $w$ or $o$ for water and oil, respectively. The evolution of saturations is governed by mass and momentum conservation laws expressed through the following partial differential equation (PDE):
\begin{equation}
    \label{eq:mass_conservation}
    \frac{\partial \phi s_a \rho_a}{\partial t} - \sum_{\gamma = 1}^{3}  \frac{\partial}{\partial x^{\gamma}} \bigg ( \frac{\rho_a k k_a}{\mu_a} \frac{\partial P} {\partial x^{\gamma}} \bigg ) = Q_{a}
\end{equation}
where $\gamma = 1, 2, 3$ is a spatial index of the coordinate vector $\mathbf{x}$, $\rho_a = \rho_a(\mathbf{x})$ and $\mu_a = \mu_a(\mathbf{x})$ are the density and viscosity of fluid $a$ at the point $\mathbf{x}$ respectively, $k = k(\mathbf{x})$ is the permeability, $\phi = \phi(\mathbf{x})$ is the porosity at a given point, $P(\textbf{x})$ is a pressure at point $\mathbf{x}$, $k_a(s)$ is a relative phase permeability of fluid $a$, $s = s(\mathbf{x})$ saturations of fluids at the point $\mathbf{x}$, $Q_a = Q_a(\mathbf{x})$ is a source term for fluid $a$ at the point $\mathbf{x}$. Generally, the permeability $k$ is a tensor. In the present example, we assume $k$ to be a spherical tensor which can vary in space. Therefore, it is fully described by a single spatial function $k = k(\mathbf{x})$. In the present work we neglect capillary pressure effects. Therefore, both fluids are subjected to the same pressure at any given point. The source terms $Q_a$ are considered to be non-zero only for cells with injection and production wells. Finally, incompressible fluids and rocks (solid matrix) are considered. Therefore, ~Eq.~\eqref{eq:mass_conservation} could be simplified:
\begin{equation}
    \label{eq:mass_conservation_volumetric}
    \phi \frac{\partial s_a}{\partial t} - \sum_{\gamma = 1}^{3} \frac{\partial}{\partial x^{\gamma}} \bigg ( \frac{k k_a}{\mu_a} \frac{\partial P} {\partial x^{\gamma}} \bigg ) = q_{a}
\end{equation}
where $q_a = Q_a/\rho_a$ is the source term for fluid $a$ normalized to the density of corresponding fluid. For calculation of relative phase permeabilities, Brooks-Corey model~\citep{Relative_Phase_Permeability} is used:
\begin{equation}
    \label{eq:Corey_model}
    \begin{gathered}
    k_{w}(S_{\text{wn}}) = k_w^{(0)} S_{\text{wn}} ^ {p_w} \\
    k_{w}(S_{\text{wn}}) = k_o^{(0)} (1-S_{\text{wn}}) ^ {p_o}
    \end{gathered}
\end{equation}
where $k_w$ and $k_o$ are the values of relative phase permeability for water and oil, respectively and $k_w^{(0)}$ and $k_o^{(0)}$ are maximum the values of relative phase permeability for water and oil, respectively. The values $p_w$ and $p_o$ are dimensionless parameters of the model and $S_{\text{wn}}$ is the normalized water saturation defined as:
\begin{equation}
    \label{eq:normalized_water_saturation}
    S_{\text{wn}} = \frac{S-S_{\text{wir}}} {1 - S_{\text{wir}} -  S_{\text{owr}}}
\end{equation}
where $S_\text{wir}$ and $S_\text{owr}$ are irreducible water and oil saturations, respectively.

In this test case, we consider a five-spot injection pattern where an injection well is located in the center of a square surrounded by four production wells. Given the symmetry of this pattern, only one quarter of the domain is modeled with one producer and one injector located at the opposite corners of a square domain. The length of the edge of that square is set to $L=640 \text{m}$. The thickness of the reservoir is $h=10 \text{m}$. We do not consider discretization along the vertical direction and we only consider a two-dimensional flow problem. For the purposes of simplicity, incompressible immiscible fluids is considered while neglecting gravity effects. A uniform square grid is used for simulations and the dimensions of each grid-block is $10 \text{m}$ by $10 \text{m}$ by $10 \text{m}$. in other words, a $64$ by $64$ by $1$ mesh is used for discretization. The porosity of the reservoir is assumed to be constant and equal to $0.2$. Both injection and production rates are considered to be constant and equal to $10\  \text{m}^3/\text{day}$. The fluid properties and parameters of Corey model are presented in the table~\ref{tab:parameters}.

\begin{table}
\begin{center}
\begin{tabular}{ |c|c|c|c|c|c| }\hline
 $\mu_o$, cP & $\mu_w$ & $p_o$ & $p_w$ &  $k_o^{(0)}$ &  $k_w^{(0)}$ \\ \hline
 $10.0$ & $1.0$ & $2.0$ & $2.0$ & $1.0$ & $1.0$  \\ \hline
\end{tabular}
\caption{Fluid properties and parameters of the model for relative-phase permeability.}
\label{tab:parameters}
\end{center}
\end{table}

In the present work, the reservoir permeability $k(\textbf{x})$ is assumed to be a random field with a predefined distribution given the correlation between values at different points within the domain. In reservoir modeling, it is natural to assume that the values of logarithm of permeability $\log(k(\mathbf{r}))$ at different points $\mathbf{r}_1$ and $\mathbf{r}_2$ are exponentially correlated~\citep{MA20117311}:
\begin{equation}
    \label{eq:exponential_correlation}
    \langle \log(k(\mathbf{r}_1)) , \log(k(\mathbf{r}_2)) \rangle = \exp\bigg( -\frac{|\mathbf{r}_1-\mathbf{r}_2|}{L_c} \bigg )
\end{equation}
where $L_c$ is a correlation length. In the present example, the correlation length is set to $L_c = 1/4 L = 160 \textit{m}$. The utilized distribution of log-permeability allows one to implement Karhunen-Loeve expansion and express $\log(k(\mathbf{r}))$ as a linear combination of mutually independent random variables:
\begin{equation}
    \label{eq:KL_expansion}
   \log(k(\mathbf{r})) = \sum_{\alpha} \theta_{\alpha} \lambda_{\alpha} \xi_{\alpha}(\mathbf{r})
\end{equation}
where $\lambda_{\alpha}, \xi_{\alpha}(\mathbf{r})$ are the eigen-values and eigen-functions of the KL expansion, respectively. The $\theta_{\alpha}$ are random mutually independent coefficients. In the present example $\theta_{\alpha}$ are considered as input random variables for the PCE response surface. The permeability field is normalized such that a zero value of $\log(k(\mathbf{r}))$ corresponds to a permeability of $1\  \text{mD}$.

\begin{figure} \centering
  \begin{subfigure}{.22\textwidth} \centering
        \includegraphics[width=0.99\linewidth]{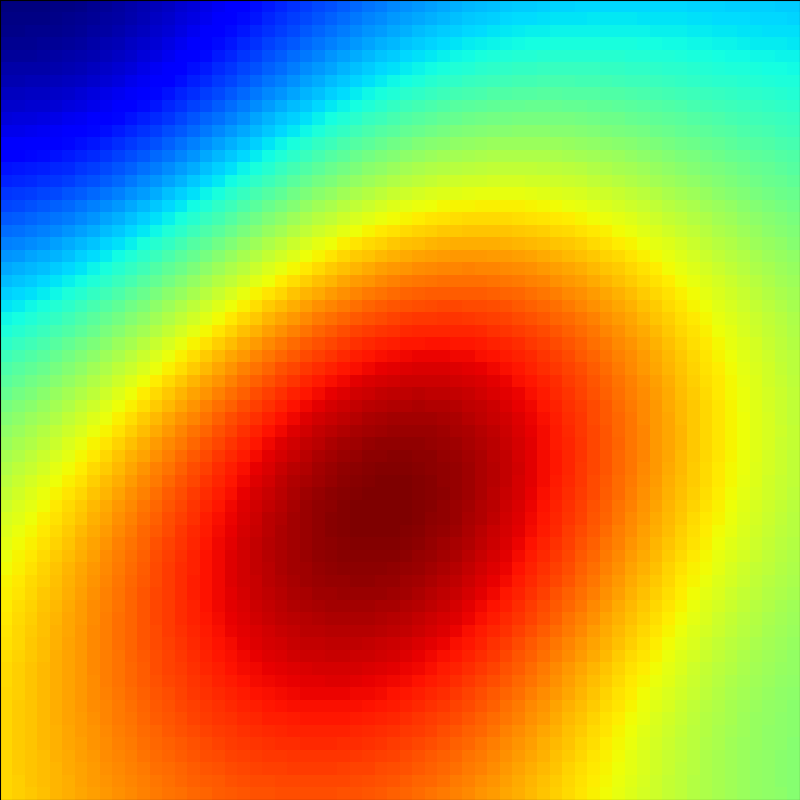}
        \caption{$5$ KL terms}
        \label{fig:log_perm_nc_5}
  \end{subfigure}
  \begin{subfigure}{.22\textwidth} \centering
        \includegraphics[width=0.99\linewidth]{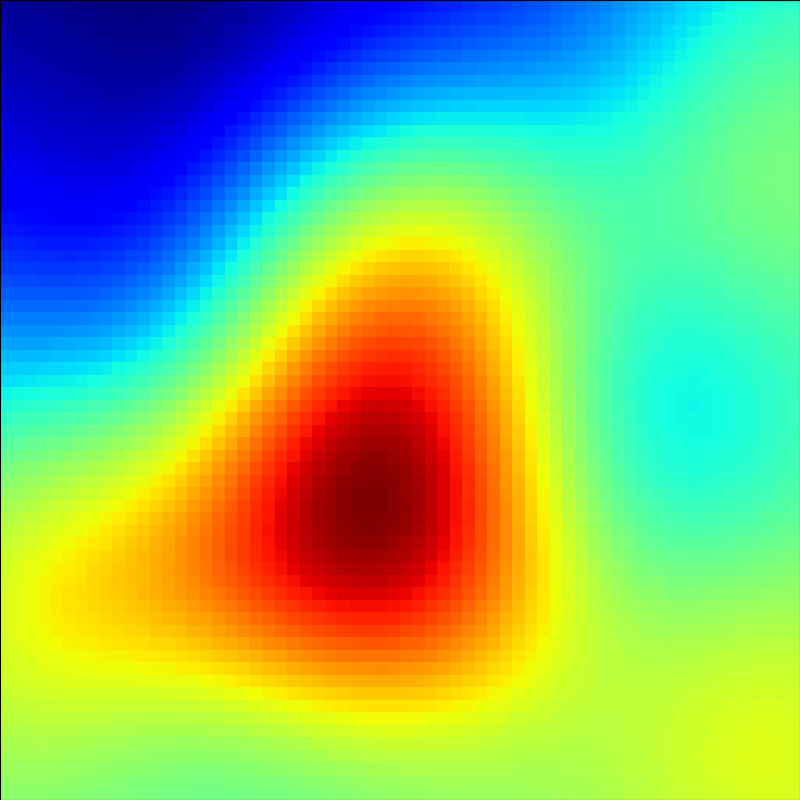}
        \caption{$15$ KL terms}
        \label{fig:log_perm_nc_15}
  \end{subfigure}
  \begin{subfigure}{.22\textwidth} \centering
        \includegraphics[width=0.99\linewidth]{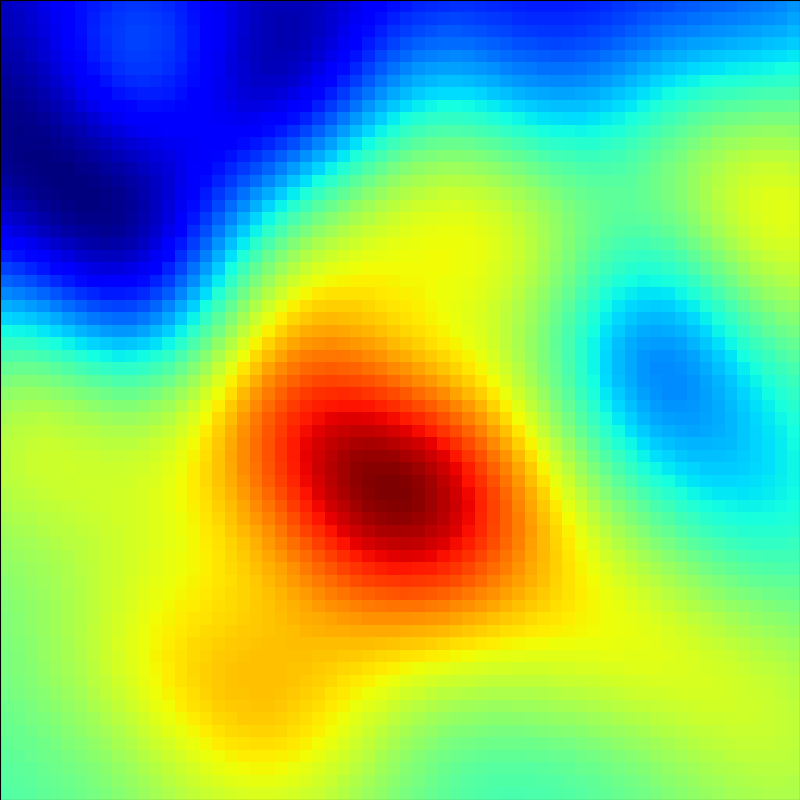}
        \caption{$45$ KL terms}
        \label{fig:log_perm_nc_45}
  \end{subfigure}
  \begin{subfigure}{.22\textwidth} \centering
        \includegraphics[width=0.99\linewidth]{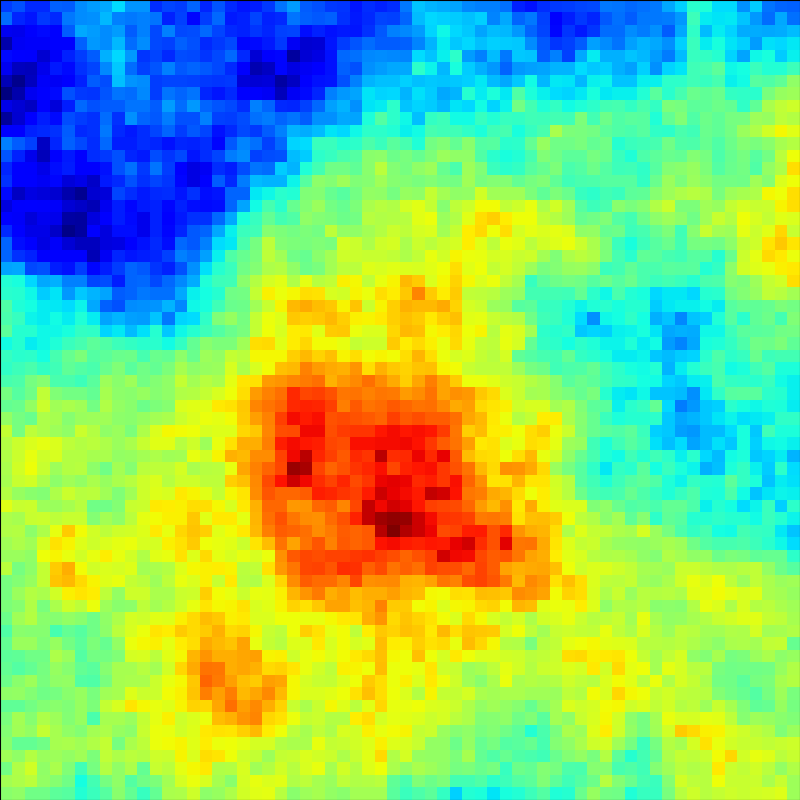}
        \caption{$4096$ KL terms}
        \label{fig:log_perm_nc_4096}
  \end{subfigure}
    \caption{Permeability realizations projected to a different number of KL-terms }
    \label{fig:log_perm}
\end{figure}

We truncate the KL expansion spectrum by taking the first $5$, $15$ or $45$ KL components. Because of the long correlation length with respect to the size of the domain, a significant part of the energy of the spectrum is captured in all truncation scenarios. In this work, the fraction of the energy of the spectrum is defined as following:
\begin{equation}
    \label{eq:energy_fraction}
    H(n) = \frac{\sum_{\alpha=1}^{n}\lambda_{\alpha}^2}{\sum_{\alpha=1}^{\infty}\lambda_{\alpha}^2}
\end{equation}
where $n$ is the number of components in the truncated KL expansion. In the present example, $H(5) = 0.9898$, $H(15) = 0.9948$ and $H(45) = 0.9972$.
It is important to notice that despite the fact that KL expansion captures significant portion of the energy spectrum, it provides smooth reconstruction of the permeability field as shown in Figure~\ref{fig:log_perm}.

Three different PCE response surfaces are built corresponding to the $5$, $15$, $45$-KL terms where $1000$ samples (i.e.~reconstructed permeability realizations) are evaluated. Each of these samples has been generated from a normal distribution of the coordinates $\theta_{\alpha}$ corresponding to the truncated eigen-vectors of the KL expansion. Water breakthrough times are estimated through numerical simulations for each of the permeability realization using a forward simulation run. A training set of $750$ samples is used for building the PCE response surface and the remaining $250$ samples are used for testing. Legendre polynomials with degree $d \leq 5$ are considered as basis functions. The tolerance in all numerical schemes used to estimate the PCE coefficients has been set to $10^{-6}$.


\begin{figure}[H]
    \centering
    \begin{subfigure}[b]{0.30\textwidth}
        \includegraphics[width=1.0\linewidth]{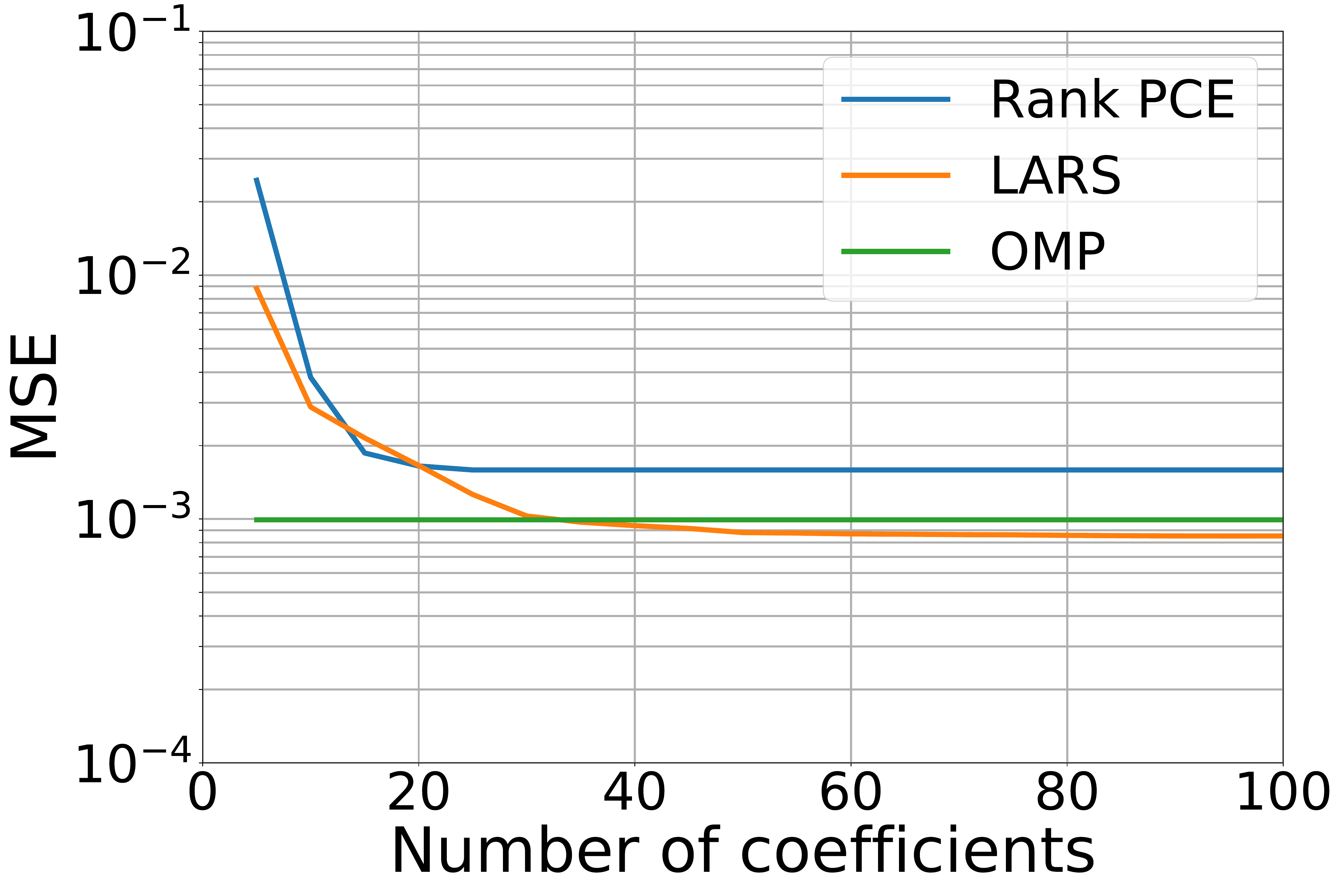}
        \caption{$5$-KL terms}
        \label{fig:mse_nkl_5}
    \end{subfigure}
    \begin{subfigure}[b]{0.30\textwidth}
        \includegraphics[width=1.0\linewidth]{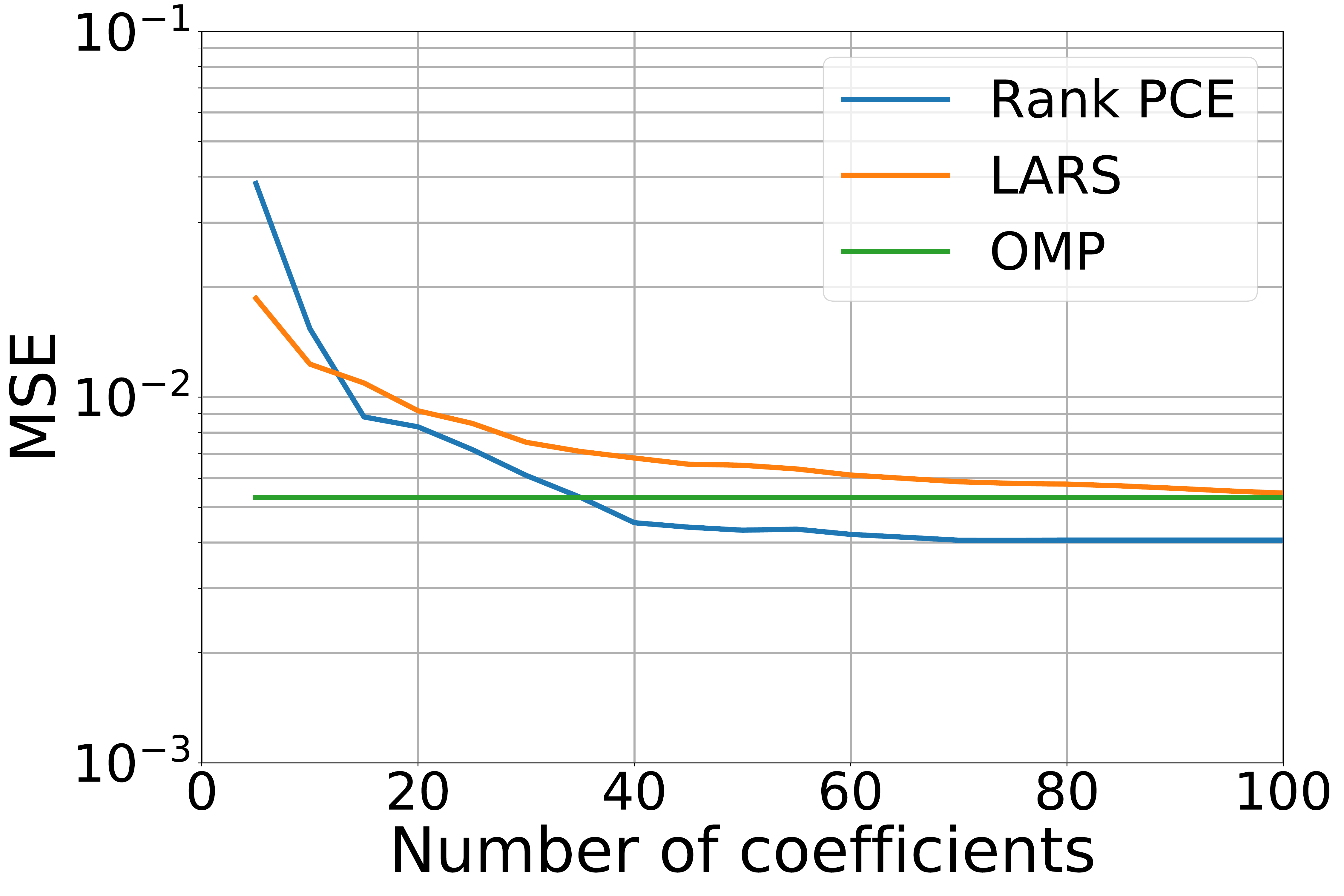}
        \caption{$15$-KL terms}
        \label{fig:mse_nkl_15}
    \end{subfigure}
    \begin{subfigure}[b]{0.30\textwidth}
        \includegraphics[width=1.0\linewidth]{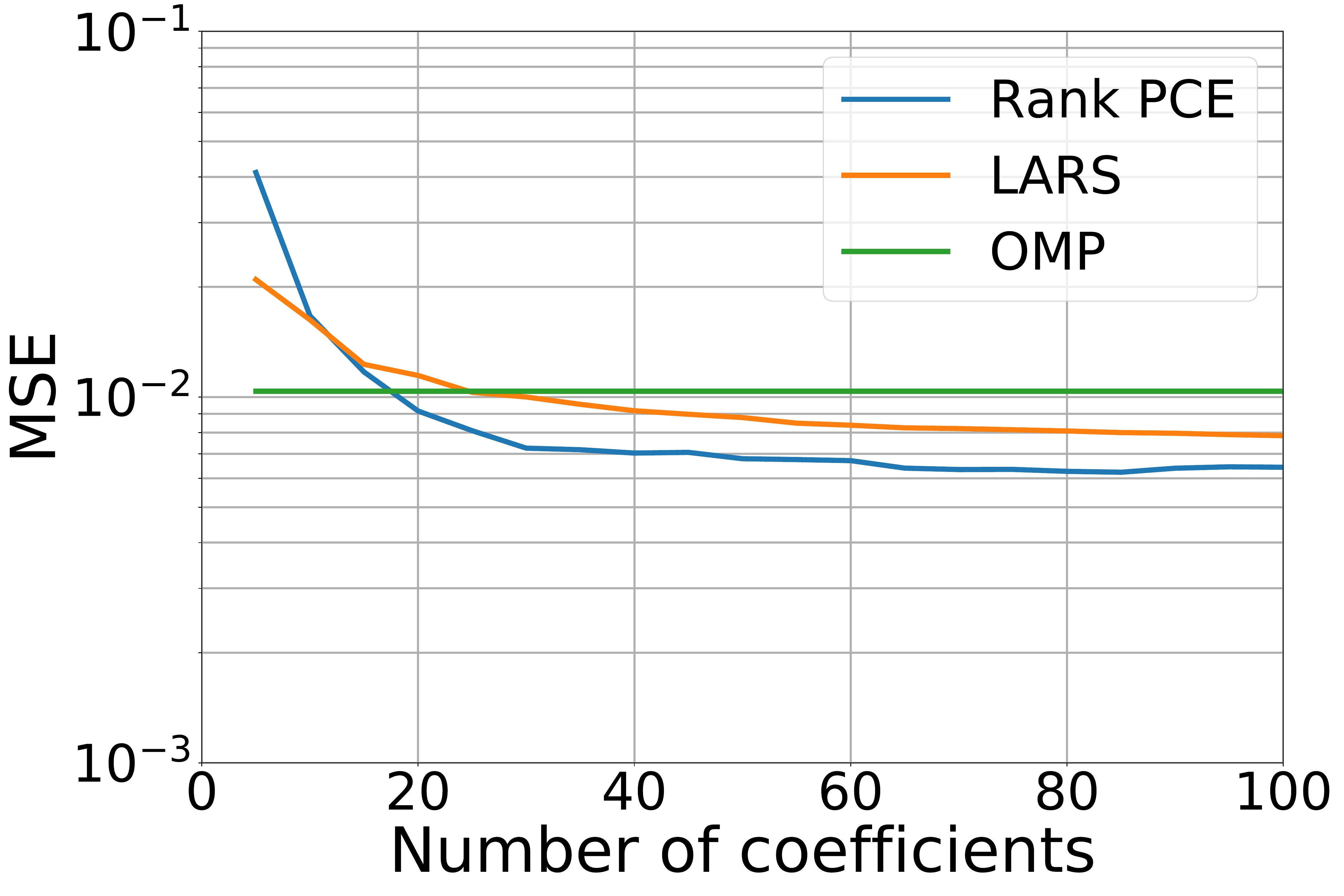}
        \caption{$45$-KL terms}
        \label{fig:mse_nkl_45}
    \end{subfigure}
    \begin{subfigure}[b]{0.30\textwidth}
        \includegraphics[width=1.0\linewidth]{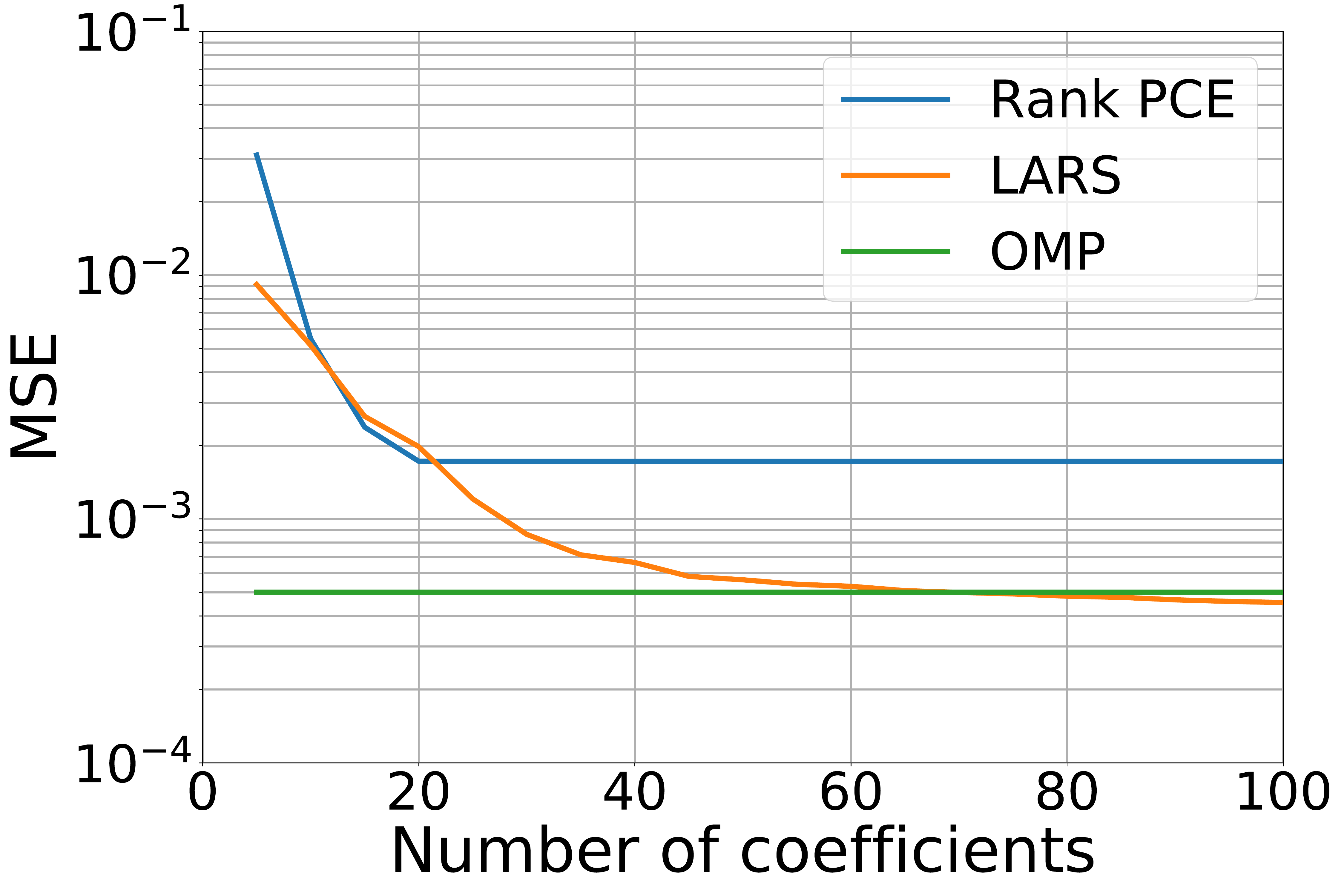}
        \caption{$5$-KL terms}
        \label{fig:mse_nkl_5_flow_rate}
    \end{subfigure}
    \begin{subfigure}[b]{0.30\textwidth}
        \includegraphics[width=1.0\linewidth]{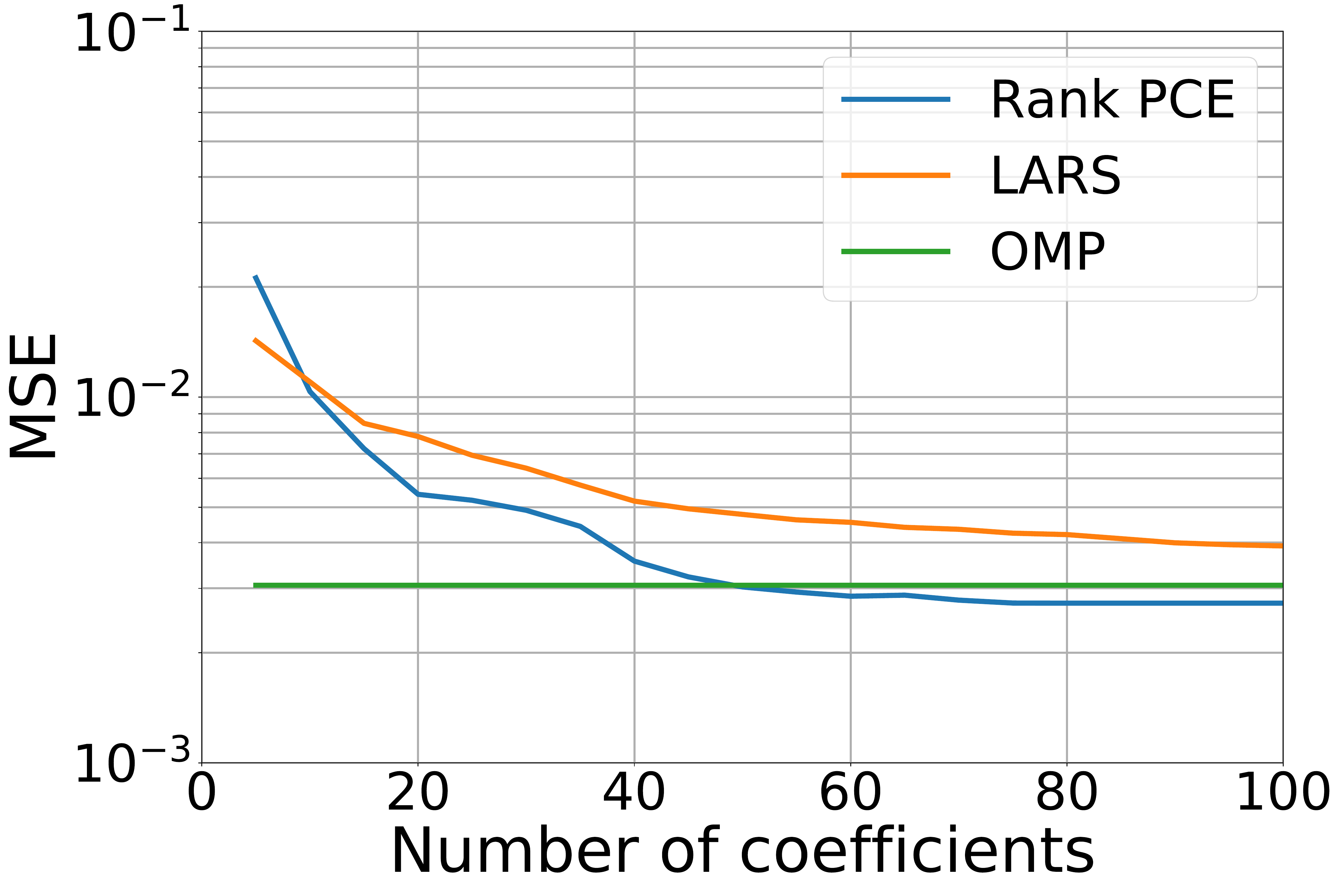}
        \caption{$15$-KL terms}
        \label{fig:mse_nkl_15_flow_rate}
    \end{subfigure}
    \begin{subfigure}[b]{0.30\textwidth}
        \includegraphics[width=1.0\linewidth]{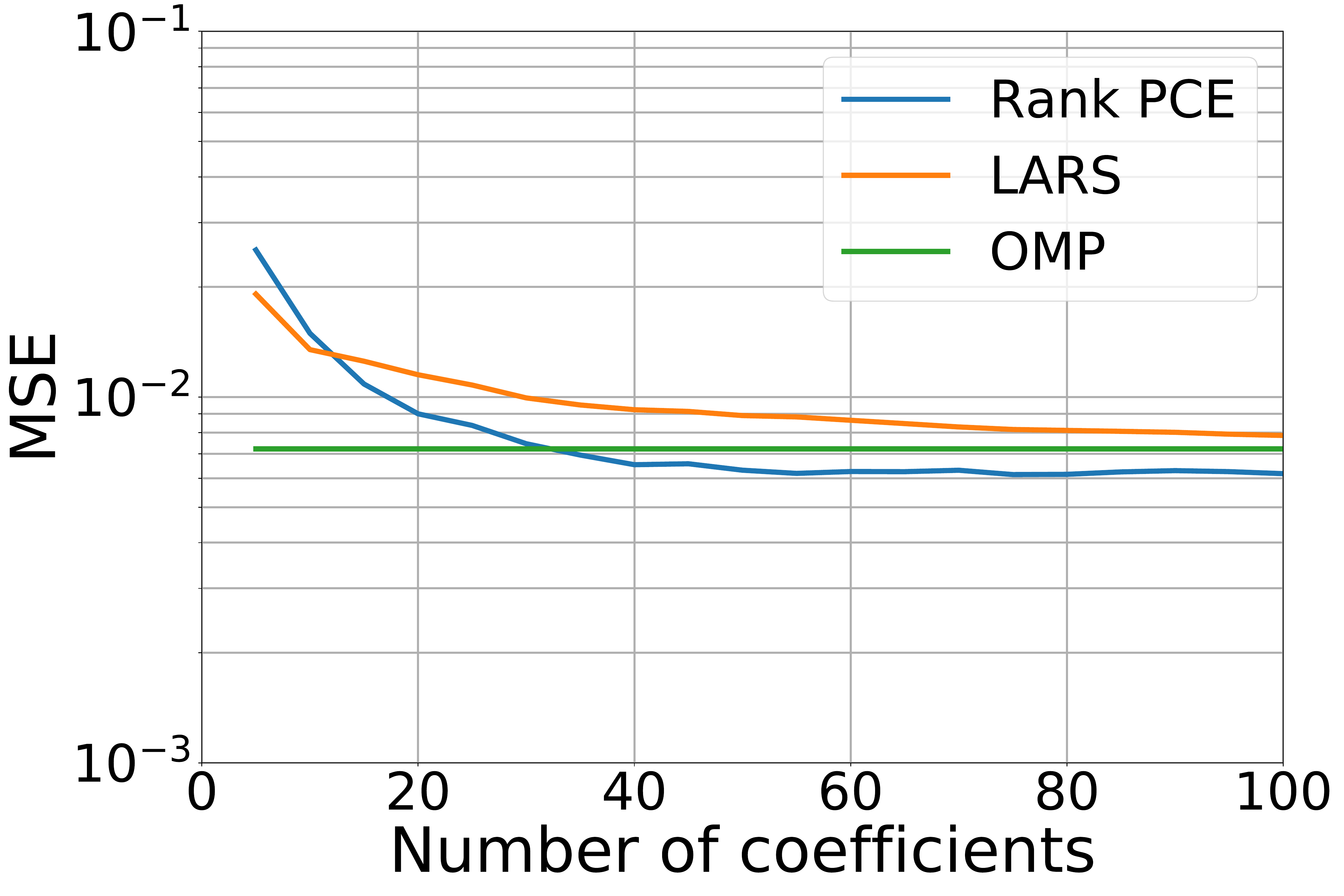}
        \caption{$45$-KL terms}
        \label{fig:mse_nkl_45_flow_rate}
    \end{subfigure}

    \caption{Mean square error (MSE) over the test data for breakthrough time (a, b, c) and for oil production rate (d, c, e) for different PCE algorithms versus the response surface free parameters.}
    \label{fig:mse_test3}

\end{figure}



\begin{figure}[H]
    \centering
    \begin{subfigure}[b]{0.32\textwidth}
        \includegraphics[width=0.99\linewidth]{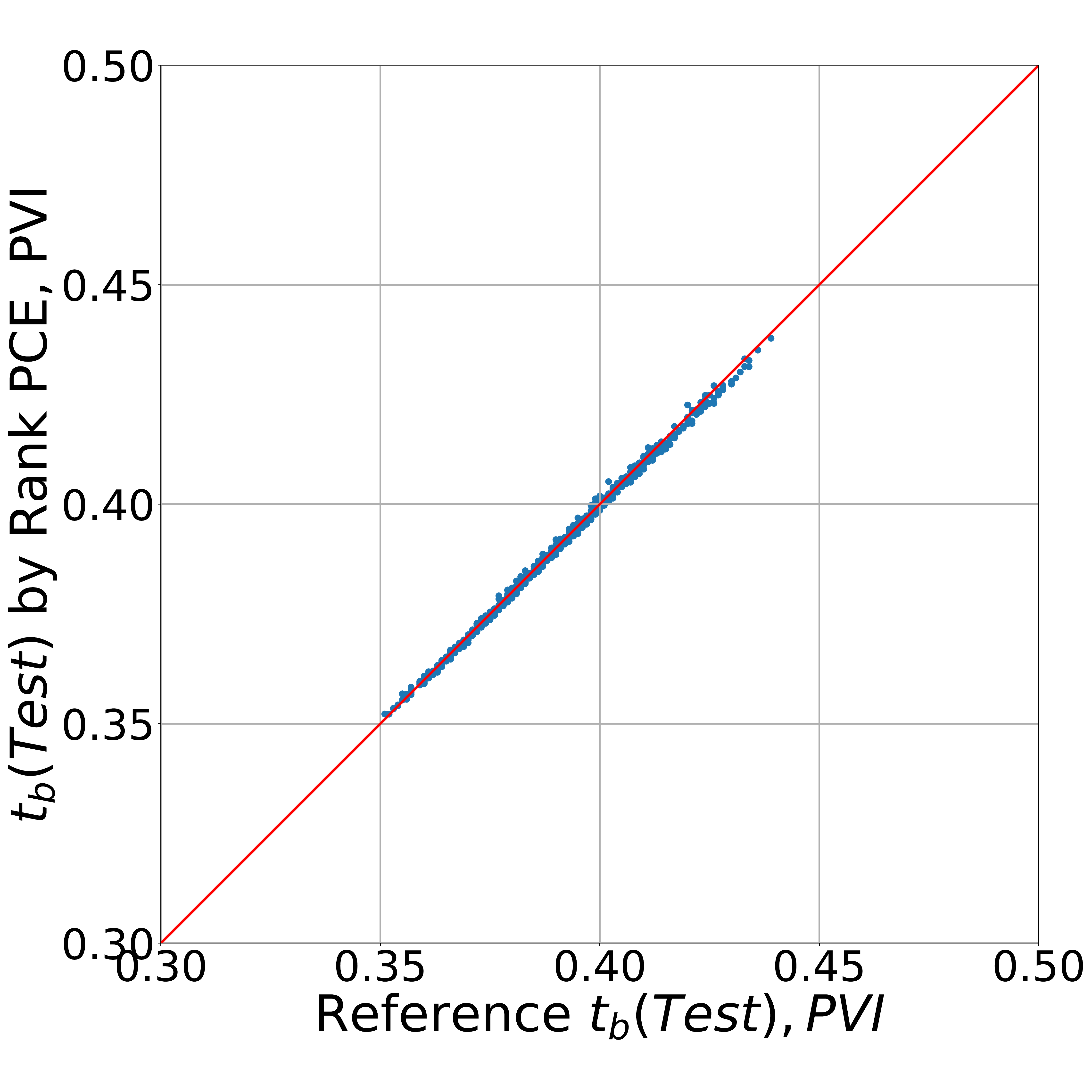}
        \caption{$5$-KL terms}
        \label{fig:cross_plot_legendre_test_nc_5}
    \end{subfigure}
    \begin{subfigure}[b]{0.32\textwidth}
        \includegraphics[width=0.99\linewidth]{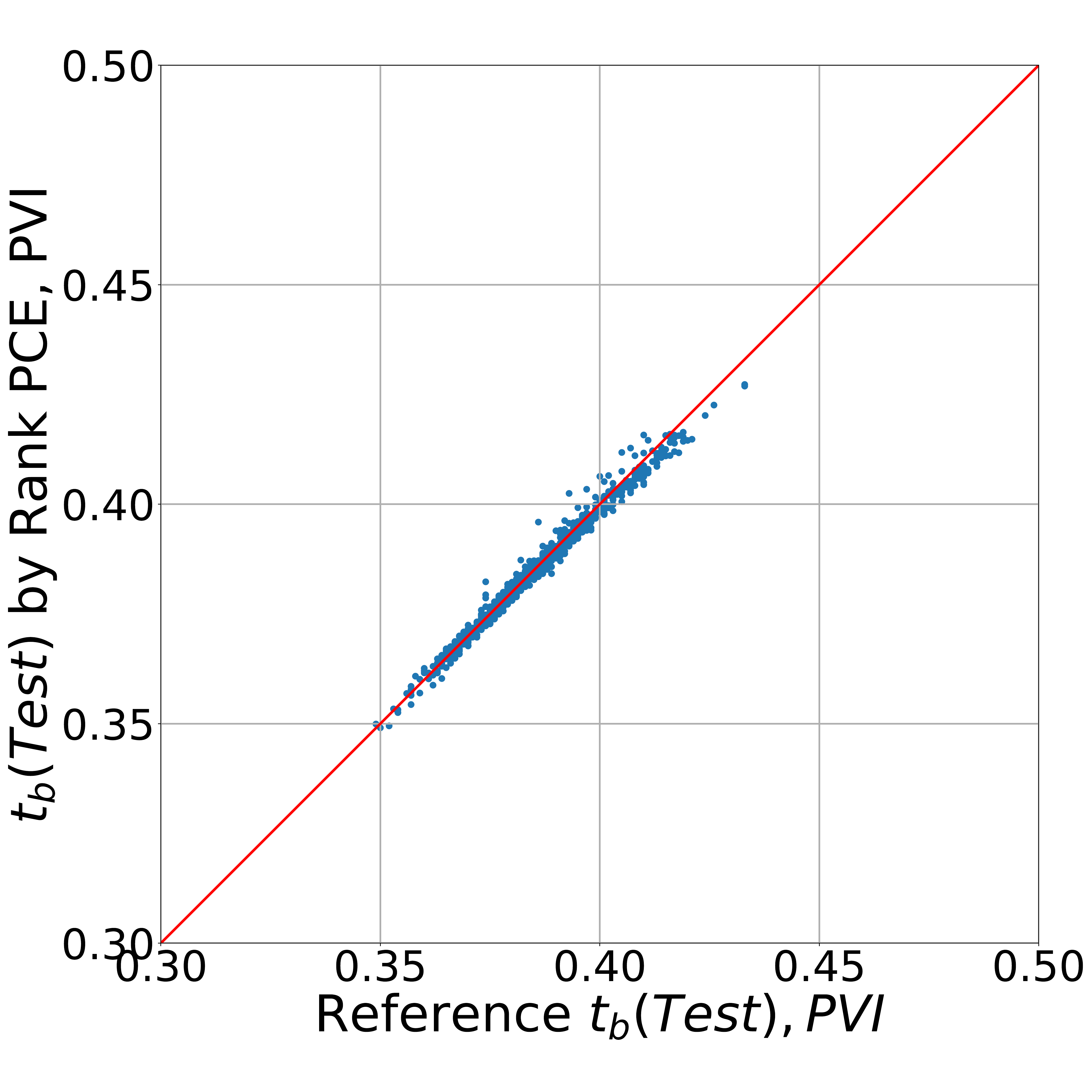}
        \caption{$15$-KL terms}
        \label{fig:cross_plot_legendre_test_nc_15}
    \end{subfigure}
    \begin{subfigure}[b]{0.32\textwidth}
        \includegraphics[width=0.99\linewidth]{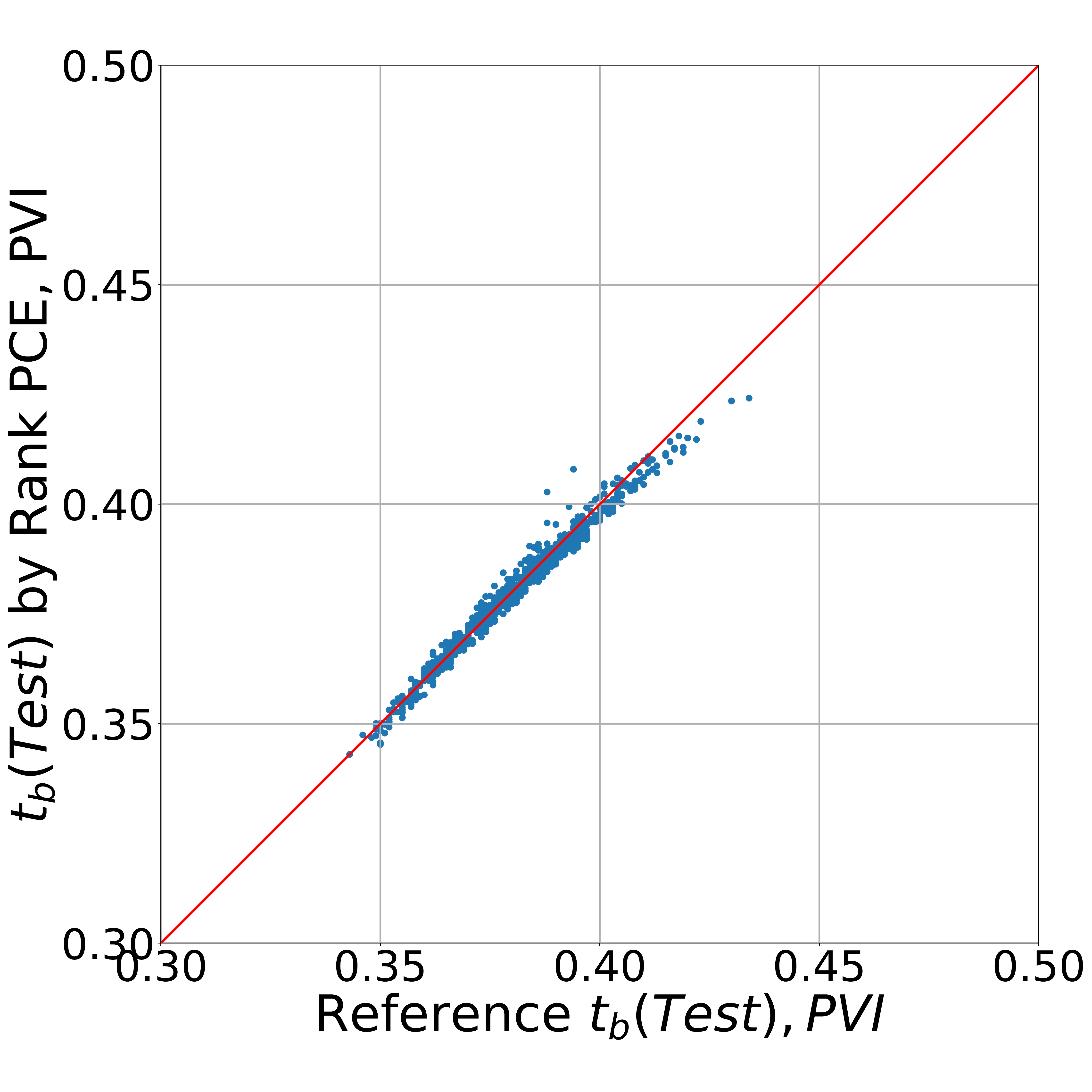}
        \caption{$45$-KL terms}
        \label{fig:cross_plot_legendre_test_nc_45}
    \end{subfigure}
    \begin{subfigure}[b]{0.32\textwidth}
        \includegraphics[width=0.99\linewidth]{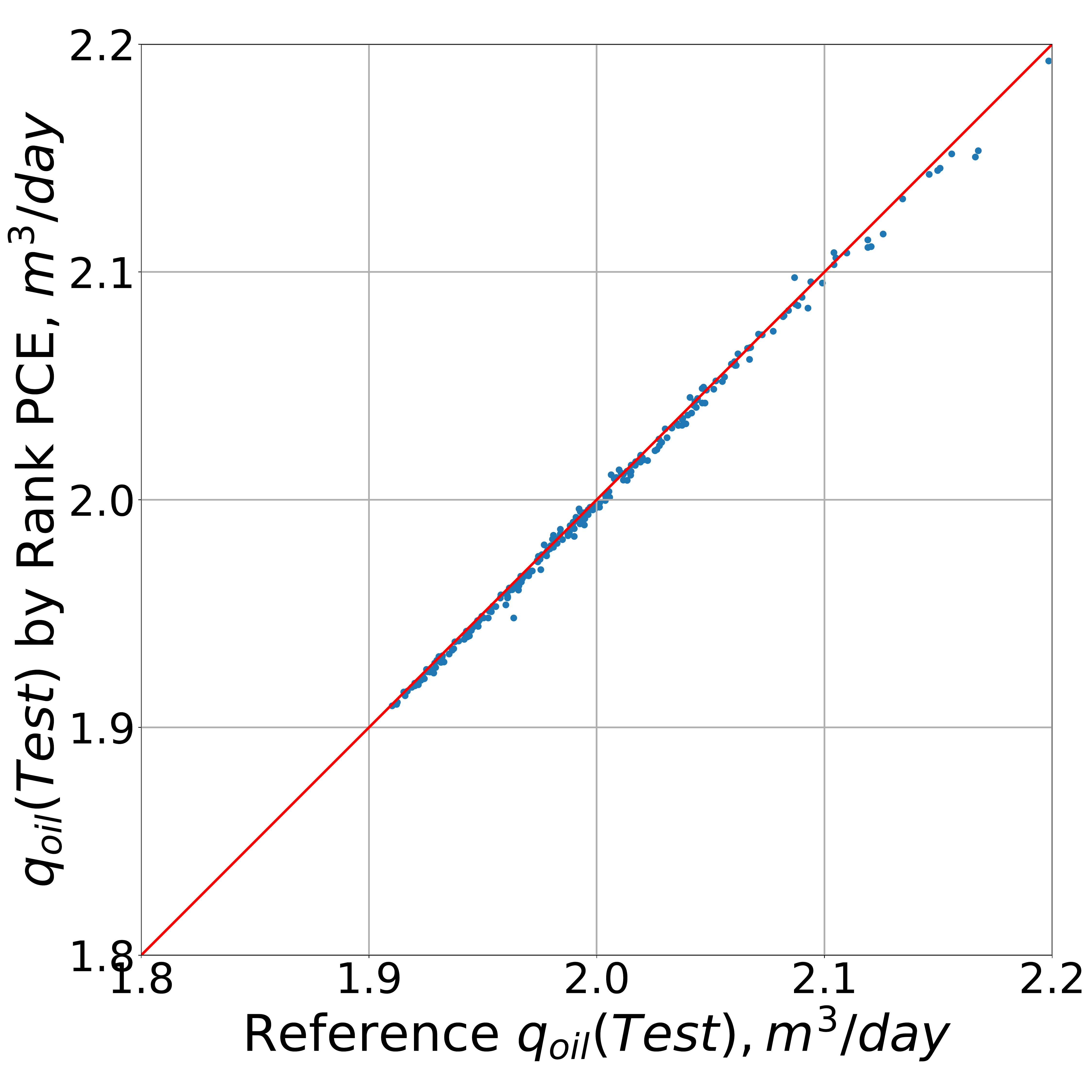}
        \caption{$5$-KL terms}
        \label{fig:cross_plot_legendre_test_nc_5_flow_rate}
    \end{subfigure}
    \begin{subfigure}[b]{0.32\textwidth}
        \includegraphics[width=0.99\linewidth]{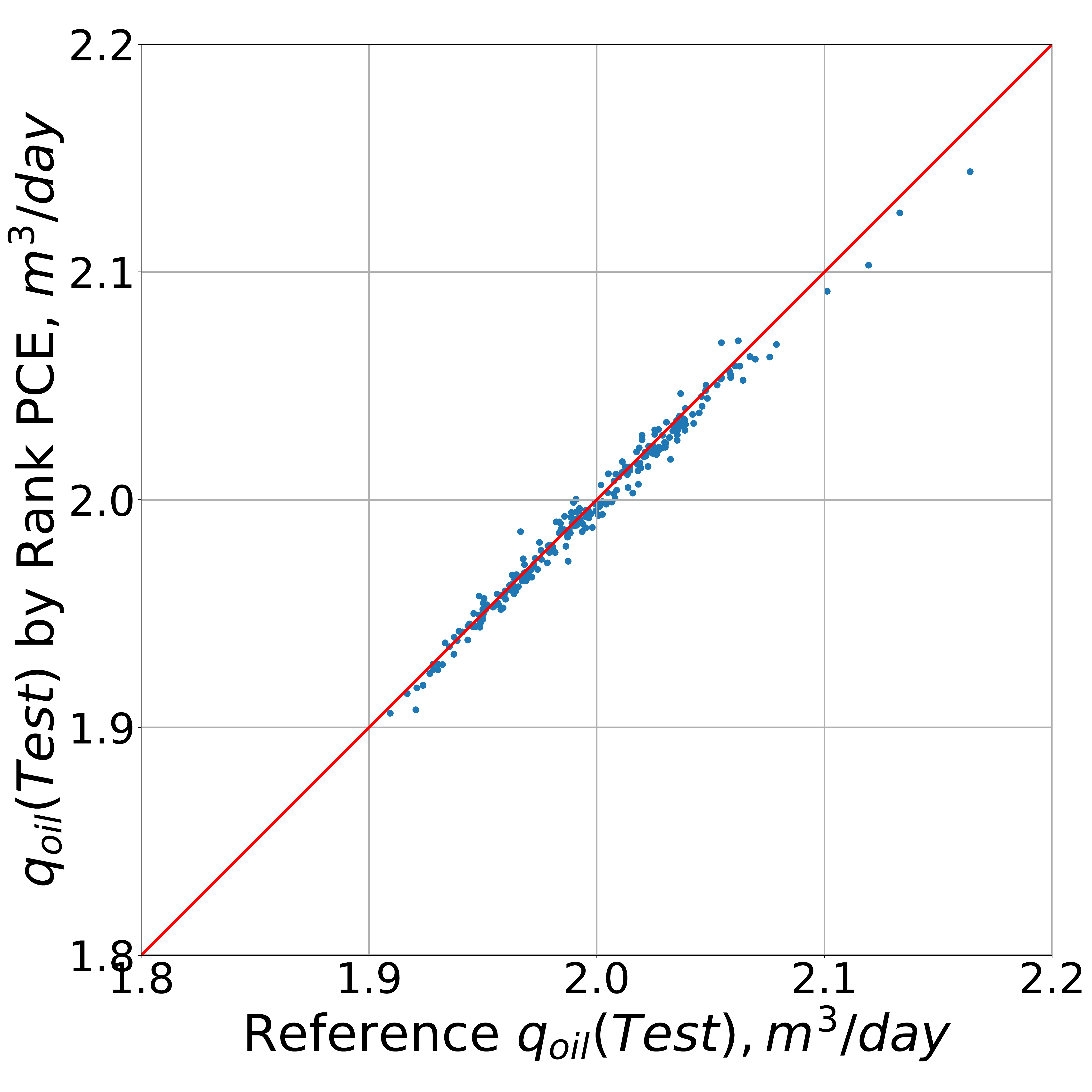}
        \caption{$15$-KL terms}
        \label{fig:cross_plot_legendre_test_nc_15_flow_rate}
    \end{subfigure}
    \begin{subfigure}[b]{0.32\textwidth}
        \includegraphics[width=0.99\linewidth]{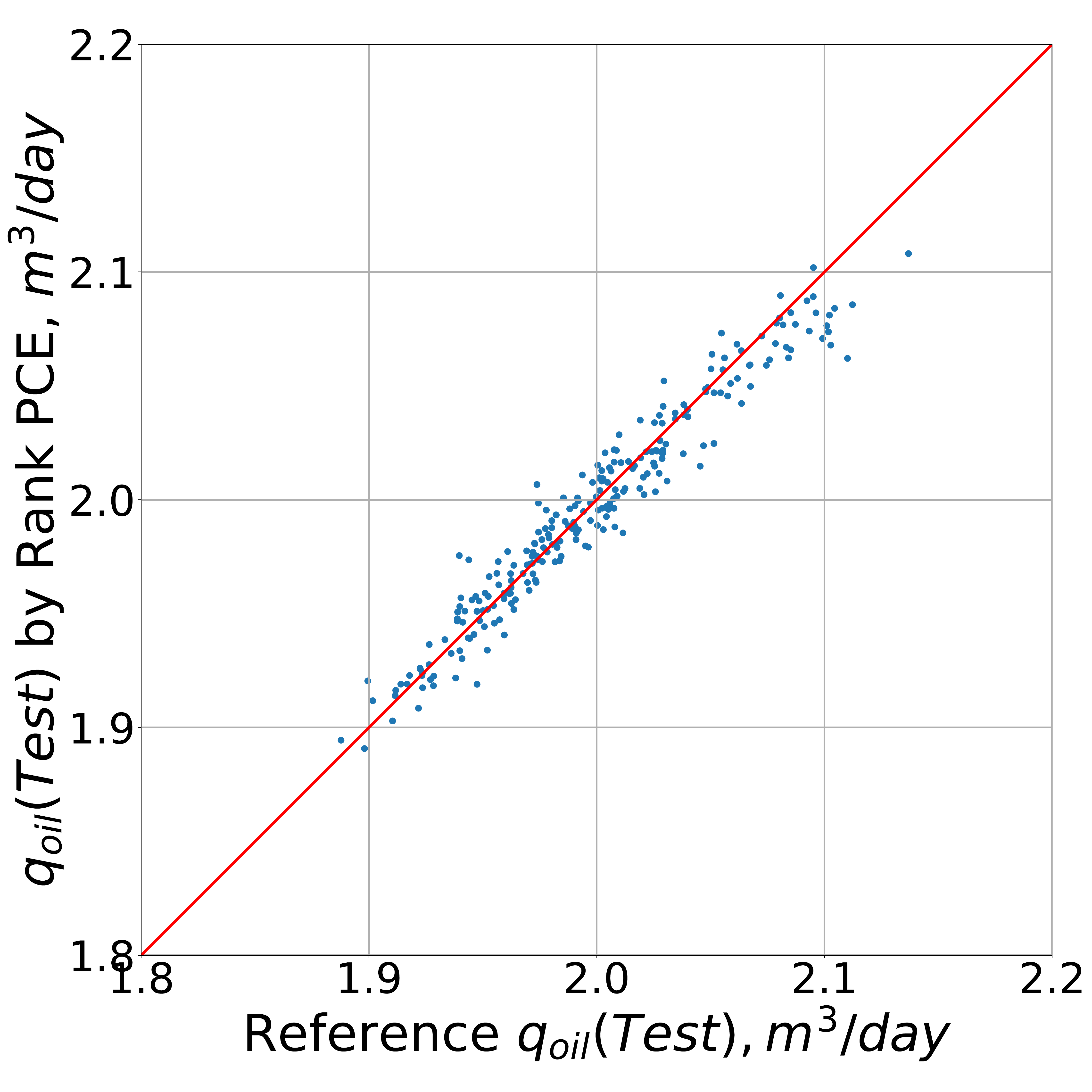}
        \caption{$45$-KL terms}
        \label{fig:cross_plot_legendre_test_nc_45_flow_rate}
    \end{subfigure}
    \caption{Cross-plots of breakthrough time and oil production rate in PVI units for Rank PCE algorithm. Top row (a, b, c) shows the results for test samples for breakthrough time and the bottom row (d, e, f) shows the results for oil production rate}
    \label{fig:cross_plot_test3}

\end{figure}

Figure~\ref{fig:mse_test3} shows the MSE for various KL truncation levels. The results presented in this figure, demonstrate that the proposed rank based PCE technique has similar accuracy when compared to the OMP algorithm for low-dimensional problems. However, the rank based PCE has clear advantages in higher dimensions. Also, both the Rank-PCE and OMP methods, perform slightly better than LARS. However, all three techniques do not perform perfectly, because the MSE is around $5\%$ of the mean-value of the QoI. The cross-plot shown in Figure~\ref{fig:cross_plot_test3} demonstrates how the quality of prediction is affected by the dimension of the problem or truncation scheme for KL expansion. The best accuracy of the response surface has been achieved for the problem with the lowest dimension corresponding to $5$-KL truncation level (left). The numerical error is the highest for $45$-KL truncation level (right). The reason for such behavior is that the training set of the same cardinality has been used for all truncation schemes. It should be noted that the accuracy of the permeability representation increases with the increase of parameter space dimension. However, capabilities of the response surface to reproduce the simulation results for a fixed number of direct simulations drop with dimension. In other words, more training data is required to build a high quality response surface for the high-dimensional case when compared with problems of lower dimensionality.

\subsection*{Test case 4: Data from CO2 injection simulations}
In this test case, PCE-based response surface is used as a fast emulator for CO2 injection process. The QoI is the mass of CO2 in a gas phase after given time period from the end of CO2 injection~\citep{UQ_CO2_2}. The data is based on the simulations results developed by~\citet{UQ_CO2_2}. The key uncertain parameters in this simulations are: average field porosity, average field permeability, regional hydraulic gradient relative phase permeability, capillary pressure and the permeability anisotropy $k_v/k_h$ ratio. More detailed description of this problem can be found in~\citep{UQ_CO2_2}. The average field porosity $\phi$ and permeability $k$ are considered as independent continuous variables with a uniform probability distribution via density-function variable transformation~\citep{rosenblatt1952}:
\begin{equation}
    \label{eq:variable_transformation}
    \begin{cases}
    \phi = f_{\phi}(x_1) \\
    k = f_{k}(x_2) \\
    \end{cases}
\end{equation}
where $x_1$ and $x_2$ are independent random variables uniformly distributed in the interval $[-1;1]$, $f_{\phi}$ and $f_{k}$ are functions for transformation of variables. All other variables are considered as discrete variables with equal probabilities over all discretized values. Table~\ref{tab:variables} summarizes the variable names and types used in this test case.

\begin{table}
\begin{center}
    \begin{tabular}{ |l|c|l| }\hline
    Variable & Notation & Type  \\ \hline
    Porosity & $x_1$ & Continuous, $\U[-1;1]$ \\ \hline
    Permeability & $x_2$ & Continuous, $\U[-1;1]$ \\ \hline
    Relative phase permeability & $x_3$ & Discrete, $10$ different models \\ \hline
    Regional hydraulic gradient & $x_4$ & Discrete,  $2$ different values\\ \hline
    Capillary pressure & $x_5$ & Discrete, $2$ different models\\ \hline
    Permeability anisotropy & $x_6$ & Discrete, $3$ different values \\ \hline
\end{tabular}
\caption{Summary of variables notations and types.}
\label{tab:variables}
\end{center}
\end{table}

\begin{figure}[H]
    \centering
    \begin{subfigure}[b]{0.45\textwidth}
        \centering
        \includegraphics[width=\textwidth]{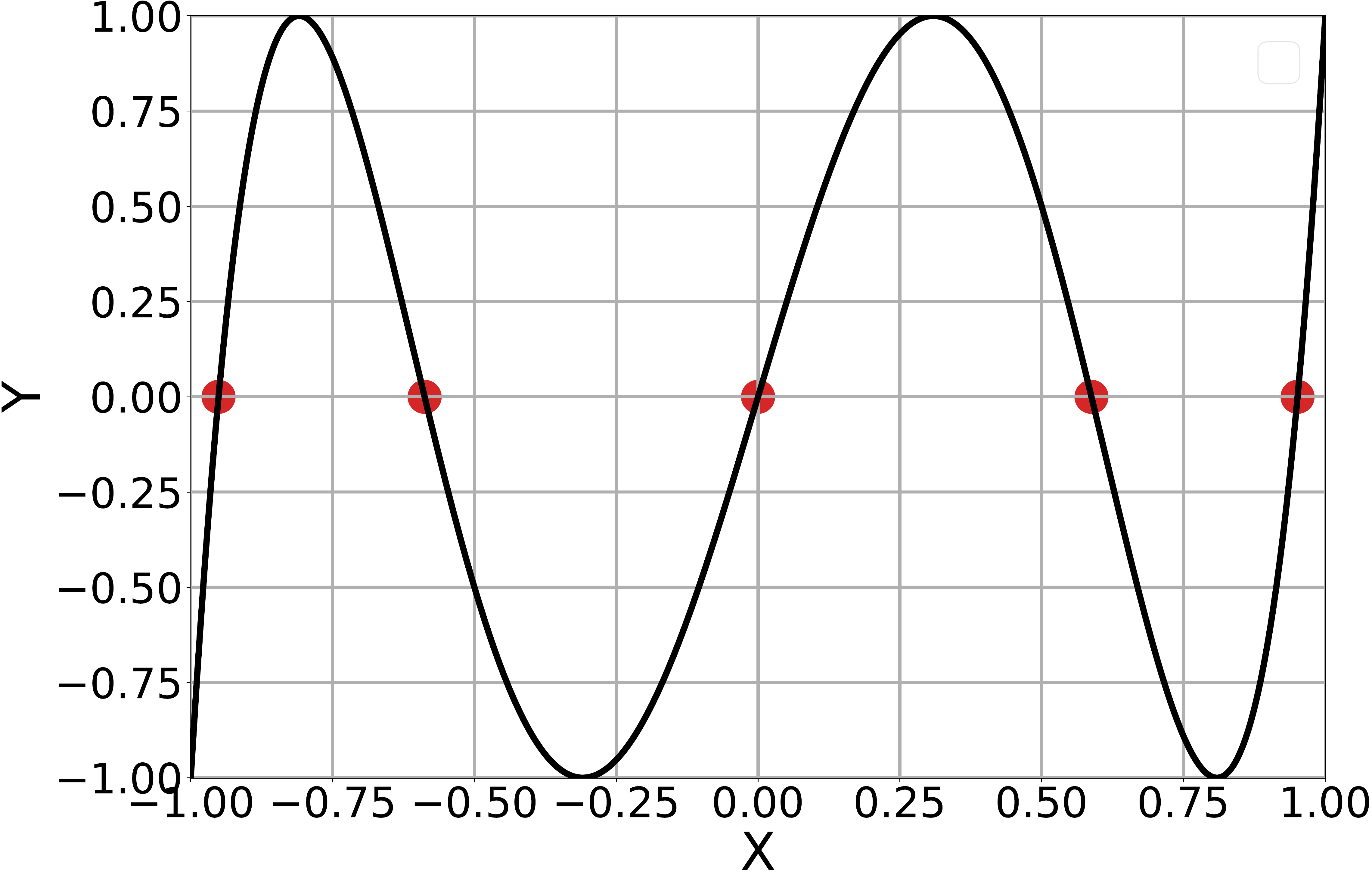}
        \caption{five distinct values}
        \label{fig:Chebyshev nodes 1}
    \end{subfigure}
    \begin{subfigure}[b]{0.45\textwidth}
        \centering
        \includegraphics[width=\textwidth]{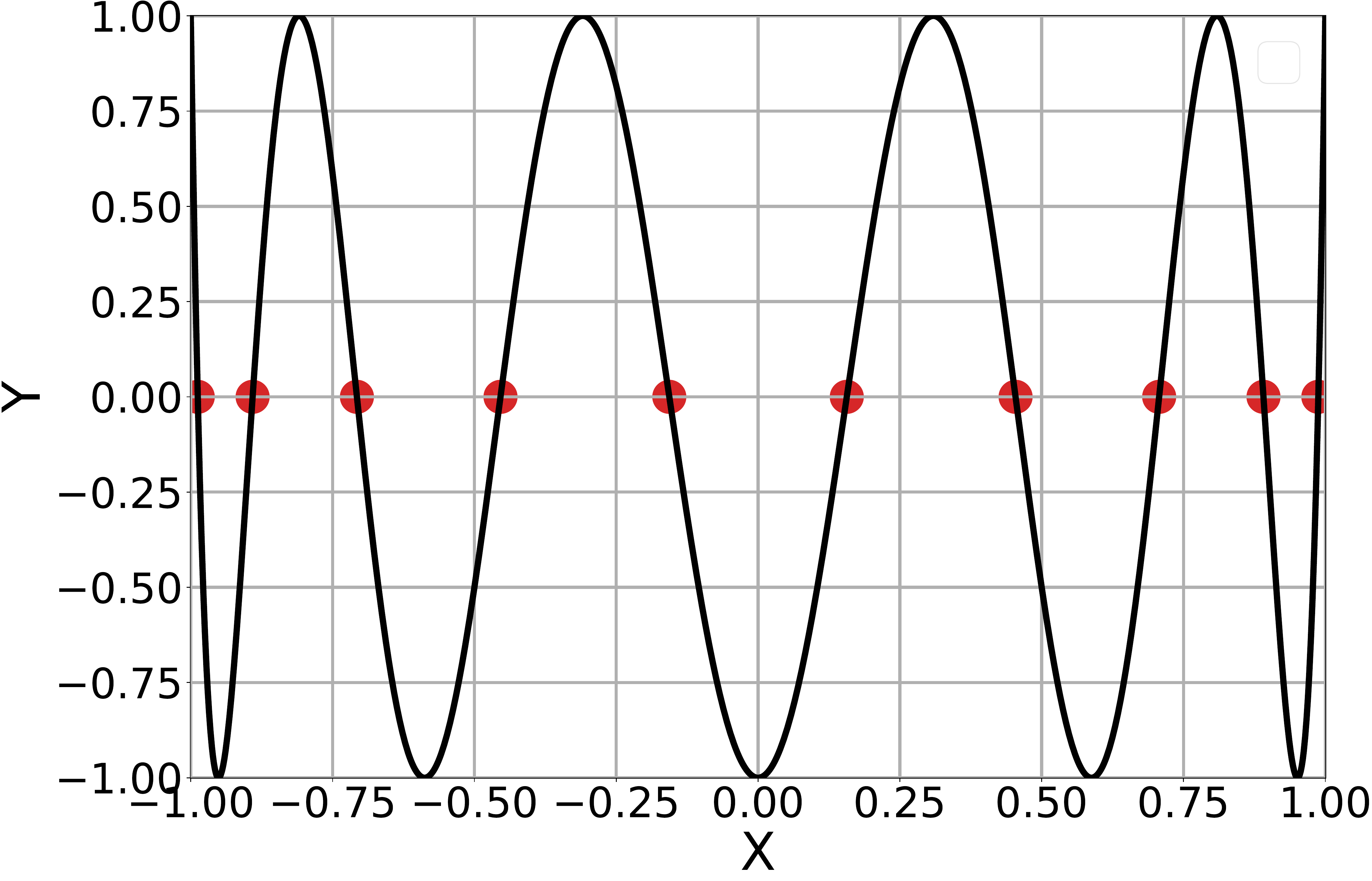}
        \caption{ten distinct values}
        \label{fig:Chebyshev nodes 2}
    \end{subfigure}
    \caption{Location of Chebyshev nodes corresponding to roots of the polynomials with the same degrees as the number of distinct values present in the categorical variable.}
    \label{fig:Chebyshev nodes}
\end{figure}

We note that this test case includes categorical variables in the input space. In order to handle this type of data, we utilize Chebyshev polynomials for categorical data. Additionally, we establish a one-to-one correspondence between the values of a given categorical variable and Chebyshev nodes:
\begin{equation}
    \label{eq:Chebyshev_nodes}
    t_m \rightarrow \cos \bigg ( \frac{2m-1}{2M}\pi\bigg)
\end{equation}
where $M$ is the total number of possible values for a given categorical variable. The mapping given by this equation is illustrated in Figure~\ref{fig:Chebyshev nodes}.

We note that Gauss-quadrature rules for Chebyshev polynomials have the same weight~\citep{Book_on_Orthogonal_Polynomials} for each of the nodes. This justifies using Chebyshev polynomials for categorical data and the corresponding mapping to the Chebyshev nodes presented in Eq.~\eqref{eq:Chebyshev_nodes} especially when training samples are uniformly distributed over the distinct categories. Therefore, the polynomials orthogonality and the distribution of categorical variables are consistent with each other.
\begin{equation}
    \label{eq:orthogonality_of_Chebyshev_polynomilas}
    \int_{-1}^{+1} \frac{ p_{\alpha}(t) p_{\beta}(t)}{ \sqrt{1-t^2}} \mathrm{d}t = \sum_m \frac{\pi }{M} p_{\alpha}(t_m) p_{\beta}(t_m) = \pi \E[p_{\alpha}p_{\beta}]
\end{equation}
In the present work we utilize normalized polynomials $p_{\alpha}(t)$ to $q_{\alpha}(t)$:
\begin{equation}
    \E[q_{\alpha}q_{\beta}] = \delta_{\alpha\beta}
\end{equation}
Using Chebyshev polynomials provides a natural extension of standard PCE to problems with categorical variables while preserving the fundamental relation between the orthogonality of basis functions and probability distribution as defined in Eq.~\eqref{eq:generic_inner_product_and_probability}.

In the current example, sampling of the data is performed using uniform distributions over the parameter ranges. A total of $998$ data points are generated in accordance with the proposed probability distributions of variables and we used $250$ data-points for training (i.e. constructing the PCE) and the remaining data points are used for testing. The mass of CO2 injection is computed via detailed numerical simulations (see~\citep{UQ_CO2_2} for more details). We normalized the QoI such that following equality holds for the training data:
\begin{equation}
    \label{eq:normalization_of_training_data}
    \sum_i \frac{y_i^2}{N} = 1
\end{equation}
We observed empirically that the QoI is highly sensitive to the permeability and relative phase permeability. Therefore, we constructed two evaluation cases with the same data set. For the first case which we refer to as the reduced case, we built a two-dimensional response surface using the permeability and relative phase permeability only as an input. The second case, which we denote as the full case, we utilize all the six uncertain variables in the response surface. In both cases, we evaluate the proposed ranking based sparse PCE against standard sparse regression PCE algorithms (i.e. LARS and OMP methods) for different numbers of expansion coefficients $N_D$. For both the reduced and full problems, PCE is performed with polynomials of degree $d \leq 10$. The number of terms in PCE varies from $5$ to $50$ and the tolerance has been set similar to all other test cases to $10^{-6}$. Legendre polynomials were used for continuous variables $x_1, x_2$ and  Chebyshev polynomials were used for the discrete/categorical variables.

\begin{figure}[H]
    \centering
    \begin{subfigure}[b]{0.45\textwidth}
        \includegraphics[width=1.0\linewidth]{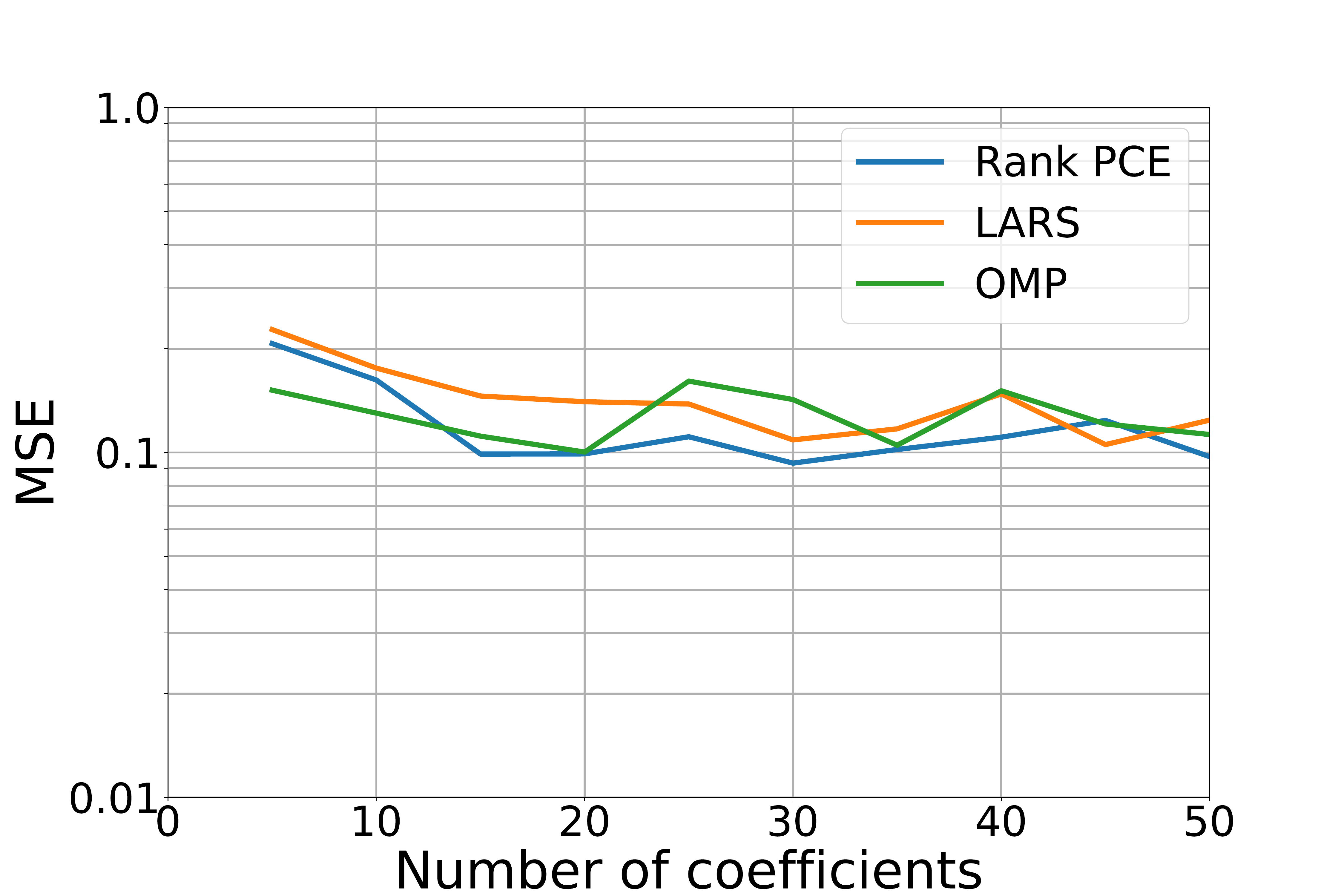}
        \caption{Reduced case with $2$ variables}
        \label{fig:mass_CO2_2d_Convergence}
    \end{subfigure}
    \begin{subfigure}[b]{0.45\textwidth}
        \includegraphics[width=1.0\linewidth]{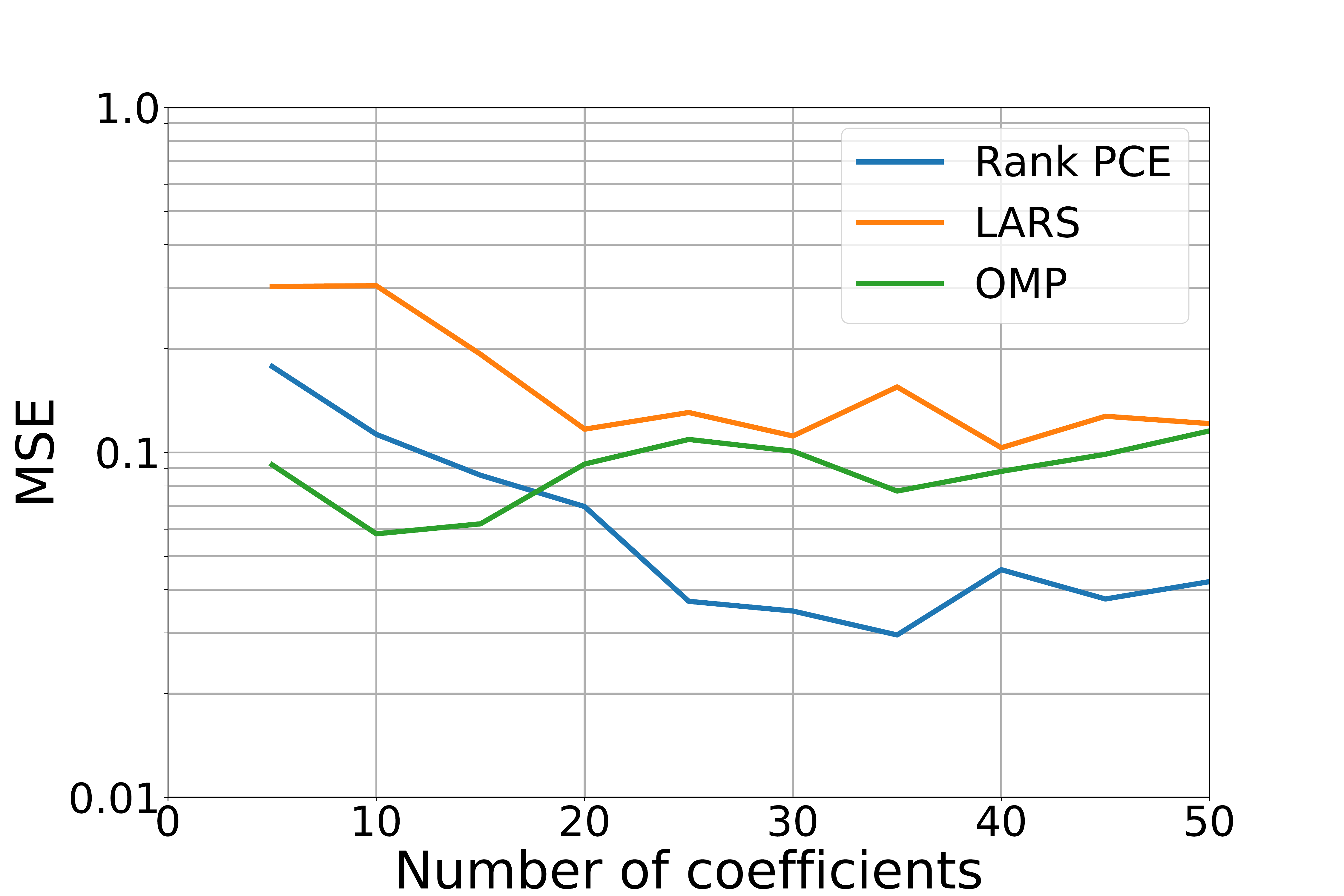}
        \caption{Full case with $6$ variables}
        \label{fig:mass_CO2_6d_Convergence}
    \end{subfigure}
    \caption{Mean square error (test data) for the different PCE algorithm versus the response surface free parameters.}
        \label{fig:mass_CO2}
\end{figure}

Figure~\ref{fig:mass_CO2}, shows the mean square error over the test data set for both the reduced and full cases in Fig.~\ref{fig:mass_CO2_2d_Convergence} and Fig.~\ref{fig:mass_CO2_6d_Convergence}, respectively. The introduced ranking based approach shows better convergence rates for both problems.
Moreover, the results in Fig.~\ref{fig:mass_CO2_6d_Convergence} demonstrate that advantages of the proposed Rank-PCE are more pronounced for higher dimensional problems, where the search space inside the iterative solver is large. For this case, the introduced ranking step allows for an efficient identification of the most significant components of PCE resulting in a higher quality response surfaces.

\section{Concluding remarks} \label{sec:conclusions}

In the current manuscript, we introduced a ranking based sparse PCE technique (Rank-PCE). The core idea of the proposed approach is to rank the PCE features in accordance with the magnitude of a given PCE coefficient based on the correlation with data while estimating for the accuracy of computed correlations. We demonstrated, via a set of numerical examples, the superior performance of Rank-PCE when compared to standard sparse regularization techniques. Rank-PCE resulted in an increase in convergence rates for generative function with sparse spectrum. We also noticed that the improvements in convergence is more pronounced for high-dimensional problems, enabling the application of PCE to problems with significant number of independent variables. Moreover, the advantages of Rank-PCE are also evident for problems with limited number of training samples as demonstrated in the analytical test cases.

In addition to novel ranking procedure, we presented an extension of PCE response surfaces to problems with both continuous and categorical data through the utilization of Chebyshev polynomials to represent the discrete variables. The proposed technique might be not optimal for general cases, however under the uniform sampling conditions, it provides a simple approach to handle categorical data in PCE that is consistent with the statistical properties of PCE for sensitivity analysis and UQ. In other words, the proposed approach maintains the relation between basis orthogonality and statistics of the input variables, which is fundamental for UQ with PCE. This technique is also easy to implement given the availability of Chebyshev polynomials in most scientific computing libraries.

\section*{Acknowledgments}

This work is funded by the European Union’s Horizon 2020 research and innovation programme under grant agreement No 653718. We thank J.C. Manceau and J. Rohmer (BRGM, France) for providing the CO2 injection dataset used in test cases. We also thank Dr. Shing Chan (formally at HWU and currently at the Oxford Big Data Institute) for developing the reservoir simulation codes used in the third example.
\newpage

\bibliographystyle{unsrtnat}
\bibliography{sample}

\begin{thebibliography}{79}
\providecommand{\natexlab}[1]{#1}
\providecommand{\url}[1]{\texttt{#1}}
\expandafter\ifx\csname urlstyle\endcsname\relax
  \providecommand{\doi}[1]{doi: #1}\else
  \providecommand{\doi}{doi: \begingroup \urlstyle{rm}\Url}\fi

\bibitem[Pedregosa et~al.(2011)Pedregosa, Varoquaux, Gramfort, Michel, Thirion,
  Grisel, Blondel, Prettenhofer, Weiss, Dubourg, Vanderplas, Passos,
  Cournapeau, Brucher, Perrot, and Duchesnay]{Scikit_Learn}
F.~Pedregosa, G.~Varoquaux, A.~Gramfort, V.~Michel, B.~Thirion, O.~Grisel,
  M.~Blondel, P.~Prettenhofer, R.~Weiss, V.~Dubourg, J.~Vanderplas, A.~Passos,
  D.~Cournapeau, M.~Brucher, M.~Perrot, and E.~Duchesnay.
\newblock Scikit-learn: Machine learning in {P}ython.
\newblock \emph{Journal of Machine Learning Research}, 12:\penalty0 2825--2830,
  2011.

\bibitem[Tugan and Sinayuc(2018)]{TUGAN2018107}
Murat~Fatih Tugan and Caglar Sinayuc.
\newblock A new fully probabilistic methodology and a software for assessing
  uncertainties and managing risks in shale gas projects at any maturity stage.
\newblock \emph{Journal of Petroleum Science and Engineering}, 168:\penalty0
  107 -- 118, 2018.
\newblock ISSN 0920-4105.
\newblock \doi{https://doi.org/10.1016/j.petrol.2018.05.001}.
\newblock URL
  \url{http://www.sciencedirect.com/science/article/pii/S0920410518303905}.

\bibitem[Wang et~al.(2019)Wang, Zhang, and Hu]{WANG2019105}
Qianlin Wang, Laibin Zhang, and Jinqiu Hu.
\newblock An integrated method of human error likelihood assessment for
  shale-gas fracturing operations based on {SPA} and {UAHP}.
\newblock \emph{Process Safety and Environmental Protection}, 123:\penalty0 105
  -- 115, 2019.
\newblock ISSN 0957-5820.
\newblock \doi{https://doi.org/10.1016/j.psep.2019.01.003}.
\newblock URL
  \url{http://www.sciencedirect.com/science/article/pii/S0957582018308449}.

\bibitem[Chen et~al.(2017)Chen, He, Wen, Chen, and Reynolds]{CHEN2017328}
Bailian Chen, Jincong He, Xian-Huan Wen, Wen Chen, and Albert~C. Reynolds.
\newblock Uncertainty quantification and value of information assessment using
  proxies and {M}arkov chain {M}onte {C}arlo method for a pilot project.
\newblock \emph{Journal of Petroleum Science and Engineering}, 157:\penalty0
  328 -- 339, 2017.
\newblock ISSN 0920-4105.
\newblock \doi{https://doi.org/10.1016/j.petrol.2017.07.039}.
\newblock URL
  \url{http://www.sciencedirect.com/science/article/pii/S0920410517305909}.

\bibitem[Ma and Zabaras(2011{\natexlab{a}})]{MA20114696}
Xiang Ma and Nicholas Zabaras.
\newblock A stochastic mixed finite element heterogeneous multiscale method for
  flow in porous media.
\newblock \emph{Journal of Computational Physics}, 230\penalty0 (12):\penalty0
  4696 -- 4722, 2011{\natexlab{a}}.
\newblock ISSN 0021-9991.
\newblock \doi{https://doi.org/10.1016/j.jcp.2011.03.001}.
\newblock URL
  \url{http://www.sciencedirect.com/science/article/pii/S0021999111001318}.

\bibitem[Zhang et~al.(2009)Zhang, Oldenburg, Finsterle, Jordan, and
  Zhang]{UQ_CO2_1}
Yingqi Zhang, Curtis~M. Oldenburg, Stefan Finsterle, Preston Jordan, and Keni
  Zhang.
\newblock Probability estimation of {CO2} leakage through faults at geologic
  carbon sequestration sites.
\newblock \emph{Energy Procedia}, 1\penalty0 (1):\penalty0 41 -- 46, 2009.
\newblock ISSN 1876-6102.
\newblock \doi{https://doi.org/10.1016/j.egypro.2009.01.008}.
\newblock URL
  \url{http://www.sciencedirect.com/science/article/pii/S1876610209000095}.
\newblock Greenhouse Gas Control Technologies 9.

\bibitem[Elgsaeter et~al.(2008)Elgsaeter, Slupphaug, and
  Johansen]{ELGSAETER20084540}
Steinar~M. Elgsaeter, Olav Slupphaug, and Tor~Arne Johansen.
\newblock Oil and gas production optimization; lost potential due to
  uncertainty.
\newblock \emph{IFAC Proceedings Volumes}, 41\penalty0 (2):\penalty0 4540 --
  4547, 2008.
\newblock ISSN 1474-6670.
\newblock \doi{https://doi.org/10.3182/20080706-5-KR-1001.00764}.
\newblock URL
  \url{http://www.sciencedirect.com/science/article/pii/S1474667016396598}.
\newblock 17th IFAC World Congress.

\bibitem[Jia et~al.(2018)Jia, McPherson, Pan, Dai, and Xiao]{JIA2018104}
Wei Jia, Brian McPherson, Feng Pan, Zhenxue Dai, and Ting Xiao.
\newblock Uncertainty quantification of {CO2} storage using bayesian model
  averaging and polynomial chaos expansion.
\newblock \emph{International Journal of Greenhouse Gas Control}, 71:\penalty0
  104 -- 115, 2018.
\newblock ISSN 1750-5836.
\newblock \doi{https://doi.org/10.1016/j.ijggc.2018.02.015}.
\newblock URL
  \url{http://www.sciencedirect.com/science/article/pii/S175058361730419X}.

\bibitem[Lackner(2003)]{Lackner1677}
Klaus~S. Lackner.
\newblock A guide to {CO2} sequestration.
\newblock \emph{Science}, 300\penalty0 (5626):\penalty0 1677--1678, 2003.
\newblock ISSN 0036-8075.
\newblock \doi{10.1126/science.1079033}.
\newblock URL \url{http://science.sciencemag.org/content/300/5626/1677}.

\bibitem[Bachu et~al.(2007)Bachu, Bonijoly, Bradshaw, Burruss, Holloway,
  Christensen, and Mathiassen]{CO2_factors}
Stefan Bachu, Didier Bonijoly, John Bradshaw, Robert Burruss, Sam Holloway,
  Niels~Peter Christensen, and Odd~Magne Mathiassen.
\newblock {CO2} storage capacity estimation: Methodology and gaps.
\newblock \emph{International Journal of Greenhouse Gas Control}, 1\penalty0
  (4):\penalty0 430 -- 443, 2007.
\newblock ISSN 1750-5836.
\newblock \doi{https://doi.org/10.1016/S1750-5836(07)00086-2}.
\newblock URL
  \url{http://www.sciencedirect.com/science/article/pii/S1750583607000862}.

\bibitem[Margaret A.~Oliver(2015)]{Book_on_Geostatistics}
Richard~Webster Margaret A.~Oliver.
\newblock \emph{{B}asic {S}teps in {G}eostatistics: The Variogram and Kriging}.
\newblock Springer, Cham, 2015.
\newblock \doi{https://doi-org.ezproxy1.hw.ac.uk/10.1007/978-3-319-15865-5}.

\bibitem[Zheng et~al.(2014)Zheng, Leung, Sawatzky, and Alvarez]{DAI2019519}
Jingwen Zheng, Juliana~Y. Leung, Ronald~P. Sawatzky, and Jose~M. Alvarez.
\newblock A cluster-based approach for visualizing and quantifying the
  uncertainty in the impacts of uncertain shale barrier configurations on
  {SAGD} production.
\newblock In \emph{SPE Canada Heavy Oil Technical Conference. Calgary, Alberta,
  Canada}, 2014.
\newblock \doi{https://doi.org/10.2118/189753-MS}.

\bibitem[Dodwell et~al.(2015)Dodwell, Ketelsen, Scheichl, and Teckentrup]{UQ2}
T.~Dodwell, C.~Ketelsen, R.~Scheichl, and A.~Teckentrup.
\newblock A hierarchical multilevel {M}arkov {C}hain {M}onte {C}arlo algorithm
  with applications to uncertainty quantification in subsurface flow.
\newblock \emph{SIAM/ASA Journal on Uncertainty Quantification}, 3\penalty0
  (1):\penalty0 1075--1108, 2015.
\newblock \doi{10.1137/130915005}.
\newblock URL \url{https://doi.org/10.1137/130915005}.

\bibitem[Khasanov et~al.(2014)Khasanov, Babin, Melchaeva, Ushmaev, Echeverria,
  and Semenikhin]{Optimization}
M.~Khasanov, V.~Babin, O.~Melchaeva, O.~Ushmaev, D.~Echeverria, and
  A.~Semenikhin.
\newblock Application of mathematical optimization techniques for well pattern
  selection.
\newblock In \emph{SPE Russian Oil and Gas Exploration \& Production Technical
  Conference and Exhibition. Moscow, Russia}, 2014.
\newblock \doi{https://doi.org/10.2118/171163-MS}.

\bibitem[Zhang et~al.(2018)Zhang, Wang, Sun, and Wang]{ZHANG2018484}
Na~Zhang, Yating Wang, Qian Sun, and Yuhe Wang.
\newblock Multiscale mass transfer coupling of triple-continuum and discrete
  fractures for flow simulation in fractured vuggy porous media.
\newblock \emph{International Journal of Heat and Mass Transfer}, 116:\penalty0
  484 -- 495, 2018.
\newblock ISSN 0017-9310.
\newblock \doi{https://doi.org/10.1016/j.ijheatmasstransfer.2017.09.046}.
\newblock URL
  \url{http://www.sciencedirect.com/science/article/pii/S0017931017316964}.

\bibitem[Li et~al.(2018)Li, Wang, and Vasilyeva]{LI2018127}
Qiuqi Li, Yuhe Wang, and Maria Vasilyeva.
\newblock Multiscale model reduction for fluid infiltration simulation through
  dual-continuum porous media with localized uncertainties.
\newblock \emph{Journal of Computational and Applied Mathematics},
  336:\penalty0 127 -- 146, 2018.
\newblock ISSN 0377-0427.
\newblock \doi{https://doi.org/10.1016/j.cam.2017.12.040}.
\newblock URL
  \url{http://www.sciencedirect.com/science/article/pii/S0377042718300086}.

\bibitem[Akkutlu et~al.(2017)Akkutlu, Efendiev, Vasilyeva, and
  Wang]{AKKUTLU201765}
I.~Yucel Akkutlu, Yalchin Efendiev, Maria Vasilyeva, and Yuhe Wang.
\newblock Multiscale model reduction for shale gas transport in a coupled
  discrete fracture and dual-continuum porous media.
\newblock \emph{Journal of Natural Gas Science and Engineering}, 48:\penalty0
  65 -- 76, 2017.
\newblock ISSN 1875-5100.
\newblock \doi{https://doi.org/10.1016/j.jngse.2017.02.040}.
\newblock URL
  \url{http://www.sciencedirect.com/science/article/pii/S1875510017300938}.
\newblock Multiscale and Multiphysics Techniques and their Applications in
  Unconventional Gas Reservoirs.

\bibitem[M.~Rame(1992)]{Dual_Mesh1}
J.E.~Killough M.~Rame.
\newblock A new approach to flow simulation in highly heterogeneous porous
  media.
\newblock \emph{SPE Formation Evaluation}, 7, 1992.
\newblock ISSN 0885-923X.
\newblock \doi{https://doi.org/10.2118/21247-PA}.

\bibitem[Audigane and J.~Blunt(2004)]{Dual_Mesh2}
Pascal Audigane and Martin J.~Blunt.
\newblock Dual mesh method for upscaling in waterflood simulation.
\newblock \emph{Transport in Porous Media}, 55:\penalty0 71--89, 01 2004.
\newblock \doi{10.1023/B:TIPM.0000007309.48913.d2}.

\bibitem[Khoozan et~al.(2011)Khoozan, Firoozabadi, Rashtchian, and
  Ashjari]{KHOOZAN2011195}
D.~Khoozan, B.~Firoozabadi, D.~Rashtchian, and M.A. Ashjari.
\newblock Analytical dual mesh method for two-phase flow through highly
  heterogeneous porous media.
\newblock \emph{Journal of Hydrology}, 400\penalty0 (1):\penalty0 195 -- 205,
  2011.
\newblock ISSN 0022-1694.
\newblock \doi{https://doi.org/10.1016/j.jhydrol.2011.01.042}.
\newblock URL
  \url{http://www.sciencedirect.com/science/article/pii/S0022169411000679}.

\bibitem[Vasilyeva et~al.(2019)Vasilyeva, Chung, Cheung, Wang, and
  Prokopev]{VASILYEVA2019}
Maria Vasilyeva, Eric~T. Chung, Siu~Wun Cheung, Yating Wang, and Georgy
  Prokopev.
\newblock Nonlocal multicontinua upscaling for multicontinua flow problems in
  fractured porous media.
\newblock \emph{Journal of Computational and Applied Mathematics}, 2019.
\newblock ISSN 0377-0427.
\newblock \doi{https://doi.org/10.1016/j.cam.2019.01.024}.
\newblock URL
  \url{http://www.sciencedirect.com/science/article/pii/S0377042719300408}.

\bibitem[Christie(1996)]{Upscaling1}
Michael Christie.
\newblock Upscaling for reservoir simulation.
\newblock \emph{Journal of Petroleum Technology - J PETROL TECHNOL},
  48:\penalty0 1004--1010, 11 1996.
\newblock \doi{10.2118/37324-MS}.

\bibitem[Qi and Hesketh(2005)]{Upscaling2}
Dasheng Qi and Tim Hesketh.
\newblock An analysis of upscaling techniques for reservoir simulation.
\newblock \emph{Petroleum Science and Technology}, 23\penalty0 (7-8):\penalty0
  827--842, 2005.
\newblock \doi{10.1081/LFT-200033132}.
\newblock URL \url{https://doi.org/10.1081/LFT-200033132}.

\bibitem[Chan and Elsheikh(2018)]{c98b95f9037d457990cec445e43f1fdd}
Shing Chan and {Ahmed H.} Elsheikh.
\newblock A machine learning approach for efficient uncertainty quantification
  using multiscale methods.
\newblock \emph{Journal of Computational Physics}, 354:\penalty0 493--511, 2
  2018.
\newblock ISSN 0021-9991.
\newblock \doi{10.1016/j.jcp.2017.10.034}.

\bibitem[Efendiev et~al.(2012)Efendiev, Galvis, and Gildin]{Model_Reduction1}
Yalchin Efendiev, Juan Galvis, and Eduardo Gildin.
\newblock Local–global multiscale model reduction for flows in high-contrast
  heterogeneous media.
\newblock \emph{Journal of Computational Physics}, 231\penalty0 (24):\penalty0
  8100 -- 8113, 2012.
\newblock ISSN 0021-9991.
\newblock \doi{https://doi.org/10.1016/j.jcp.2012.07.032}.
\newblock URL
  \url{http://www.sciencedirect.com/science/article/pii/S0021999112004160}.

\bibitem[Carlberg et~al.(2018)Carlberg, Choi, and Sargsyan]{Model_Reduction2}
Kevin Carlberg, Youngsoo Choi, and Syuzanna Sargsyan.
\newblock Conservative model reduction for finite-volume models.
\newblock \emph{Journal of Computational Physics}, 371:\penalty0 280 -- 314,
  2018.
\newblock ISSN 0021-9991.
\newblock \doi{https://doi.org/10.1016/j.jcp.2018.05.019}.
\newblock URL
  \url{http://www.sciencedirect.com/science/article/pii/S002199911830319X}.

\bibitem[Gosses et~al.(2018)Gosses, Nowak, and Wöhling]{Model_Reduction3}
Moritz Gosses, Wolfgang Nowak, and Thomas Wöhling.
\newblock Explicit treatment for {D}irichlet, {N}eumann and {C}auchy boundary
  conditions in {POD}-based reduction of groundwater models.
\newblock \emph{Advances in Water Resources}, 115:\penalty0 160 -- 171, 2018.
\newblock ISSN 0309-1708.
\newblock \doi{https://doi.org/10.1016/j.advwatres.2018.03.011}.
\newblock URL
  \url{http://www.sciencedirect.com/science/article/pii/S0309170817307467}.

\bibitem[Kani and Elsheikh(2018)]{2ae730559d3d4dfa92f1da3e3188cc79}
{J. Nagoor} Kani and {Ahmed H.} Elsheikh.
\newblock Reduced-order modeling of subsurface multi-phase flow models using
  deep residual recurrent neural networks.
\newblock \emph{Transport in Porous Media}, pages 1--29, 10 2018.
\newblock ISSN 0169-3913.
\newblock \doi{10.1007/s11242-018-1170-7}.

\bibitem[Mo et~al.(2019)Mo, Zhu, Zabaras, Shi, and Wu]{Proxy1}
Shaoxing Mo, Yinhao Zhu, Nicholas Zabaras, Xiaoqing Shi, and Jichun Wu.
\newblock Deep convolutional encoder-decoder networks for uncertainty
  quantification of dynamic multiphase flow in heterogeneous media.
\newblock \emph{Water Resources Research}, 55\penalty0 (1):\penalty0 703--728,
  2019.
\newblock \doi{10.1029/2018WR023528}.
\newblock URL
  \url{https://agupubs.onlinelibrary.wiley.com/doi/abs/10.1029/2018WR023528}.

\bibitem[Agada et~al.(2016)Agada, Geiger, Elsheikh, and Oladyshkin]{Proxy2}
Simeon Agada, Sebastian Geiger, Ahmed Elsheikh, and Sergey Oladyshkin.
\newblock Data-driven surrogates for rapid simulation and optimization of {WAG}
  injection in fractured carbonate reservoirs.
\newblock \emph{Petroleum Geoscience}, 2016.
\newblock ISSN 1354-0793.
\newblock \doi{10.1144/petgeo2016-068}.
\newblock URL
  \url{https://pg.lyellcollection.org/content/early/2016/12/10/petgeo2016-068}.

\bibitem[Josset et~al.(2015)Josset, Demyanov, Elsheikh, and Lunati]{Proxy3}
Laureline Josset, Vasily Demyanov, Ahmed Elsheikh, and Ivan Lunati.
\newblock Accelerating {M}onte {C}arlo {M}arkov chains with proxy and error
  models.
\newblock \emph{Computers {\&} Geosciences}, 85:\penalty0 38 -- 48, 2015.
\newblock ISSN 0098-3004.
\newblock \doi{https://doi.org/10.1016/j.cageo.2015.07.003}.
\newblock URL
  \url{http://www.sciencedirect.com/science/article/pii/S009830041530011X}.
\newblock Statistical learning in geoscience modelling: Novel algorithms and
  challenging case studies.

\bibitem[Schöbi and Sudret(2018)]{PCE_Stat1}
Roland Schöbi and Bruno Sudret.
\newblock Global sensitivity analysis in the context of imprecise probabilities
  (p-boxes) using sparse polynomial chaos expansions.
\newblock \emph{Reliability Engineering \& System Safety}, 2018.
\newblock ISSN 0951-8320.
\newblock \doi{https://doi.org/10.1016/j.ress.2018.11.021}.
\newblock URL
  \url{http://www.sciencedirect.com/science/article/pii/S0951832017306099}.

\bibitem[Camacho et~al.(2017)Camacho, Talavera, Emerick, Pacheco, and
  Zanni]{PCE_quadrature}
Alejandra Camacho, Alvaro Talavera, Alexandre~A. Emerick, Marco~A.C. Pacheco,
  and João Zanni.
\newblock Uncertainty quantification in reservoir simulation models with
  polynomial chaos expansions: {S}molyak quadrature and regression method
  approach.
\newblock \emph{Journal of Petroleum Science and Engineering}, 153:\penalty0
  203 -- 211, 2017.
\newblock ISSN 0920-4105.
\newblock \doi{https://doi.org/10.1016/j.petrol.2017.03.046}.
\newblock URL
  \url{http://www.sciencedirect.com/science/article/pii/S0920410517303960}.

\bibitem[WU et~al.(2018)WU, ZHANG, SONG, and YE]{WU2018997}
Xiaojing WU, Weiwei ZHANG, Shufang SONG, and Zhengyin YE.
\newblock Sparse grid-based polynomial chaos expansion for aerodynamics of an
  airfoil with uncertainties.
\newblock \emph{Chinese Journal of Aeronautics}, 31\penalty0 (5):\penalty0 997
  -- 1011, 2018.
\newblock ISSN 1000-9361.
\newblock \doi{https://doi.org/10.1016/j.cja.2018.03.011}.
\newblock URL
  \url{http://www.sciencedirect.com/science/article/pii/S1000936118301031}.

\bibitem[Xu and Kong(2018)]{XU201824}
Jun Xu and Fan Kong.
\newblock A cubature collocation based sparse polynomial chaos expansion for
  efficient structural reliability analysis.
\newblock \emph{Structural Safety}, 74:\penalty0 24 -- 31, 2018.
\newblock ISSN 0167-4730.
\newblock \doi{https://doi.org/10.1016/j.strusafe.2018.04.001}.
\newblock URL
  \url{http://www.sciencedirect.com/science/article/pii/S0167473017303922}.

\bibitem[Palar et~al.(2018)Palar, Zuhal, Shimoyama, and Tsuchiya]{PALAR2018175}
Pramudita~Satria Palar, Lavi~Rizki Zuhal, Koji Shimoyama, and Takeshi Tsuchiya.
\newblock Global sensitivity analysis via multi-fidelity polynomial chaos
  expansion.
\newblock \emph{Reliability Engineering \& System Safety}, 170:\penalty0 175 --
  190, 2018.
\newblock ISSN 0951-8320.
\newblock \doi{https://doi.org/10.1016/j.ress.2017.10.013}.
\newblock URL
  \url{http://www.sciencedirect.com/science/article/pii/S0951832016304872}.

\bibitem[Abraham et~al.(2017)Abraham, Raisee, Ghorbaniasl, Contino, and
  Lacor]{PCE_Regression}
S.~Abraham, M.~Raisee, G.~Ghorbaniasl, F.~Contino, and C.~Lacor.
\newblock A robust and efficient stepwise regression method for building sparse
  polynomial chaos expansions.
\newblock \emph{Journal of Computational Physics}, 332:\penalty0 461 -- 474,
  2017.
\newblock ISSN 0021-9991.
\newblock \doi{https://doi.org/10.1016/j.jcp.2016.12.015}.
\newblock URL
  \url{http://www.sciencedirect.com/science/article/pii/S0021999116306684}.

\bibitem[Alemazkoor and Meidani(2018)]{PCE_Preconditioning}
Negin Alemazkoor and Hadi Meidani.
\newblock A preconditioning approach for improved estimation of sparse
  polynomial chaos expansions.
\newblock \emph{Computer Methods in Applied Mechanics and Engineering},
  342:\penalty0 474 -- 489, 2018.
\newblock ISSN 0045-7825.
\newblock \doi{https://doi.org/10.1016/j.cma.2018.08.005}.
\newblock URL
  \url{http://www.sciencedirect.com/science/article/pii/S0045782518303918}.

\bibitem[Abolghasemi et~al.(2012)Abolghasemi, Ferdowsi, and
  Sanei]{ABOLGHASEMI2012999}
Vahid Abolghasemi, Saideh Ferdowsi, and Saeid Sanei.
\newblock A gradient-based alternating minimization approach for optimization
  of the measurement matrix in compressive sensing.
\newblock \emph{Signal Processing}, 92\penalty0 (4):\penalty0 999 -- 1009,
  2012.
\newblock ISSN 0165-1684.
\newblock \doi{https://doi.org/10.1016/j.sigpro.2011.10.012}.
\newblock URL
  \url{http://www.sciencedirect.com/science/article/pii/S0165168411003665}.

\bibitem[{Li} et~al.(2013){Li}, {Zhu}, {Yang}, {Chang}, and {Bai}]{6484193}
G.~{Li}, Z.~{Zhu}, D.~{Yang}, L.~{Chang}, and H.~{Bai}.
\newblock On projection matrix optimization for compressive sensing systems.
\newblock \emph{IEEE Transactions on Signal Processing}, 61\penalty0
  (11):\penalty0 2887--2898, June 2013.
\newblock ISSN 1053-587X.
\newblock \doi{10.1109/TSP.2013.2253776}.

\bibitem[Hampton and Doostan(2015)]{PCE_Hybrid}
Jerrad Hampton and Alireza Doostan.
\newblock Coherence motivated sampling and convergence analysis of least
  squares polynomial chaos regression.
\newblock \emph{Computer Methods in Applied Mechanics and Engineering},
  290:\penalty0 73 -- 97, 2015.
\newblock ISSN 0045-7825.
\newblock \doi{https://doi.org/10.1016/j.cma.2015.02.006}.
\newblock URL
  \url{http://www.sciencedirect.com/science/article/pii/S004578251500047X}.

\bibitem[Hosder et~al.(2010)Hosder, Walters, and Balch]{Hosder_2010}
Serhat Hosder, Robert~W. Walters, and Michael Balch.
\newblock Point-collocation nonintrusive polynomial chaos method for stochastic
  computational fluid dynamics.
\newblock \emph{{AIAA} Journal}, 48\penalty0 (12):\penalty0 2721--2730, dec
  2010.
\newblock \doi{10.2514/1.39389}.
\newblock URL \url{https://doi.org/10.2514\%2F1.39389}.

\bibitem[Liang~Yan(2012)]{Yan_2012}
Dongbin~Xiu Liang~Yan, Ling~Guo.
\newblock Stochastic collocation algorithms using l1 minimization.
\newblock \emph{International Journal for Uncertainty Quantification},
  2\penalty0 (3):\penalty0 279--293, 2012.
\newblock ISSN 2152-5080.

\bibitem[Wai-Tsun~Ng and Eldred(2012)]{good_collocation}
Leo Wai-Tsun~Ng and Michael Eldred.
\newblock Multifidelity uncertainty quantification using non-intrusive
  polynomial chaos and stochastic collocation.
\newblock 04 2012.
\newblock ISBN 978-1-60086-937-2.
\newblock \doi{10.2514/6.2012-1852}.

\bibitem[Masoud~Babaei(2015{\natexlab{a}})]{Babaei_Paper1}
Ali~Alkhatib Masoud~Babaei, Indranil~Pan.
\newblock Robust optimization of well location to enhance hysteretical trapping
  of {CO2}: Assessment of various uncertainty quantification methods and
  utilization of mixed response surface surrogates.
\newblock \emph{Water Resources Research}, 51\penalty0 (12):\penalty0
  9402--9424, 2015{\natexlab{a}}.
\newblock \doi{10.1002/2015WR017418}.
\newblock URL
  \url{https://agupubs.onlinelibrary.wiley.com/doi/abs/10.1002/2015WR017418}.

\bibitem[Masoud~Babaei(2015{\natexlab{b}})]{Babaei_Paper2}
Indranil~Pan Masoud~Babaei, Ali~Alkhatib.
\newblock Robust optimization of subsurface flow using polynomial chaos and
  response surface surrogates.
\newblock \emph{Computational Geosciences}, 19\penalty0 (5):\penalty0 979--998,
  Oct 2015{\natexlab{b}}.
\newblock ISSN 1573-1499.
\newblock \doi{10.1007/s10596-015-9516-5}.
\newblock URL \url{https://doi.org/10.1007/s10596-015-9516-5}.

\bibitem[Petvipusit et~al.(2014)Petvipusit, Elsheikh, Laforce, King, and
  Blunt]{Petvipusit2014}
Kurt~R. Petvipusit, Ahmed~H. Elsheikh, Tara~C. Laforce, Peter~R. King, and
  Martin~J. Blunt.
\newblock Robust optimisation of {CO2} sequestration strategies under
  geological uncertainty using adaptive sparse grid surrogates.
\newblock \emph{Computational Geosciences}, 18\penalty0 (5):\penalty0 763--778,
  Oct 2014.
\newblock ISSN 1573-1499.
\newblock \doi{10.1007/s10596-014-9425-z}.
\newblock URL \url{https://doi.org/10.1007/s10596-014-9425-z}.

\bibitem[Elsheikh et~al.(2014)Elsheikh, Hoteit, and Wheeler]{ELSHEIKH2014515}
Ahmed~H. Elsheikh, Ibrahim Hoteit, and Mary~F. Wheeler.
\newblock Efficient bayesian inference of subsurface flow models using nested
  sampling and sparse polynomial chaos surrogates.
\newblock \emph{Computer Methods in Applied Mechanics and Engineering},
  269:\penalty0 515 -- 537, 2014.
\newblock ISSN 0045-7825.
\newblock \doi{https://doi.org/10.1016/j.cma.2013.11.001}.
\newblock URL
  \url{http://www.sciencedirect.com/science/article/pii/S004578251300296X}.

\bibitem[Petropoulos et~al.(2018)Petropoulos, Kourentzes, Nikolopoulos, and
  Siemsen]{Model_Selection}
Fotios Petropoulos, Nikolaos Kourentzes, Konstantinos Nikolopoulos, and Enno
  Siemsen.
\newblock Judgmental selection of forecasting models.
\newblock \emph{Journal of Operations Management}, 60:\penalty0 34 -- 46, 2018.
\newblock ISSN 0272-6963.
\newblock \doi{https://doi.org/10.1016/j.jom.2018.05.005}.
\newblock URL
  \url{http://www.sciencedirect.com/science/article/pii/S0272696318300251}.

\bibitem[Hampton and Doostan(2018)]{HAMPTON201820}
Jerrad Hampton and Alireza Doostan.
\newblock Basis adaptive sample efficient polynomial chaos ({BASE}-{PC}).
\newblock \emph{Journal of Computational Physics}, 371:\penalty0 20 -- 49,
  2018.
\newblock ISSN 0021-9991.
\newblock \doi{https://doi.org/10.1016/j.jcp.2018.03.035}.
\newblock URL
  \url{http://www.sciencedirect.com/science/article/pii/S0021999118301955}.

\bibitem[Bazargan et~al.(2015)Bazargan, Christie, Elsheikh, and
  Ahmadi]{BAZARGAN2015385}
Hamid Bazargan, Mike Christie, Ahmed~H. Elsheikh, and Mohammad Ahmadi.
\newblock Surrogate accelerated sampling of reservoir models with complex
  structures using sparse polynomial chaos expansion.
\newblock \emph{Advances in Water Resources}, 86:\penalty0 385 -- 399, 2015.
\newblock ISSN 0309-1708.
\newblock \doi{https://doi.org/10.1016/j.advwatres.2015.09.009}.
\newblock URL
  \url{http://www.sciencedirect.com/science/article/pii/S030917081500216X}.
\newblock Data assimilation for improved predictions of integrated terrestrial
  systems.

\bibitem[Konakli and Sudret(2016)]{Sparse_PCE_L1}
Katerina Konakli and Bruno Sudret.
\newblock Polynomial meta-models with canonical low-rank approximations:
  {N}umerical insights and comparison to sparse polynomial chaos expansions.
\newblock \emph{Journal of Computational Physics}, 321:\penalty0 1144 -- 1169,
  2016.
\newblock ISSN 0021-9991.
\newblock \doi{https://doi.org/10.1016/j.jcp.2016.06.005}.
\newblock URL
  \url{http://www.sciencedirect.com/science/article/pii/S0021999116302303}.

\bibitem[Meng and Li(2017)]{classic_sparse_PCE}
Jin Meng and Heng Li.
\newblock An efficient stochastic approach for flow in porous media via sparse
  polynomial chaos expansion constructed by feature selection.
\newblock \emph{Advances in Water Resources}, 105:\penalty0 13 -- 28, 2017.
\newblock ISSN 0309-1708.
\newblock \doi{https://doi.org/10.1016/j.advwatres.2017.04.019}.
\newblock URL
  \url{http://www.sciencedirect.com/science/article/pii/S030917081630625X}.

\bibitem[Guo et~al.(2018{\natexlab{a}})Guo, Narayan, and
  Zhou]{PCE_Derivatives2}
Ling Guo, Akil Narayan, and Tao Zhou.
\newblock A gradient enhanced $\ell$1-minimization for sparse approximation of
  polynomial chaos expansions.
\newblock \emph{Journal of Computational Physics}, 367:\penalty0 49 -- 64,
  2018{\natexlab{a}}.
\newblock ISSN 0021-9991.
\newblock \doi{https://doi.org/10.1016/j.jcp.2018.04.026}.
\newblock URL
  \url{http://www.sciencedirect.com/science/article/pii/S0021999118302420}.

\bibitem[Thapa et~al.(2018)Thapa, Mulani, and Walters]{PCE_Derivatives1}
Mishal Thapa, Sameer~B. Mulani, and Robert~W. Walters.
\newblock A new non-intrusive polynomial chaos using higher order
  sensitivities.
\newblock \emph{Computer Methods in Applied Mechanics and Engineering},
  328:\penalty0 594 -- 611, 2018.
\newblock ISSN 0045-7825.
\newblock \doi{https://doi.org/10.1016/j.cma.2017.09.024}.
\newblock URL
  \url{http://www.sciencedirect.com/science/article/pii/S0045782517306539}.

\bibitem[Cheng and Lu(2018)]{PCE_SVR}
Kai Cheng and Zhenzhou Lu.
\newblock Adaptive sparse polynomial chaos expansions for global sensitivity
  analysis based on support vector regression.
\newblock \emph{Computers \& Structures}, 194:\penalty0 86 -- 96, 2018.
\newblock ISSN 0045-7949.
\newblock \doi{https://doi.org/10.1016/j.compstruc.2017.09.002}.
\newblock URL
  \url{http://www.sciencedirect.com/science/article/pii/S0045794917305047}.

\bibitem[Pranesh and Ghosh(2018)]{PCE_PCG}
Srikara Pranesh and Debraj Ghosh.
\newblock Cost reduction of stochastic {G}alerkin method by adaptive
  identification of significant polynomial chaos bases for elliptic equations.
\newblock \emph{Computer Methods in Applied Mechanics and Engineering},
  340:\penalty0 54 -- 69, 2018.
\newblock ISSN 0045-7825.
\newblock \doi{https://doi.org/10.1016/j.cma.2018.04.043}.
\newblock URL
  \url{http://www.sciencedirect.com/science/article/pii/S0045782518302287}.

\bibitem[Guo et~al.(2018{\natexlab{b}})Guo, Dias, Carvajal, Peyras, and
  Breul]{PCE_Residual_Based_Ranking}
Xiangfeng Guo, Daniel Dias, Claudio Carvajal, Laurent Peyras, and Pierre Breul.
\newblock Reliability analysis of embankment dam sliding stability using the
  sparse polynomial chaos expansion.
\newblock \emph{Engineering Structures}, 174:\penalty0 295 -- 307,
  2018{\natexlab{b}}.
\newblock ISSN 0141-0296.
\newblock \doi{https://doi.org/10.1016/j.engstruct.2018.07.053}.
\newblock URL
  \url{http://www.sciencedirect.com/science/article/pii/S014102961830511X}.

\bibitem[Blatman and Sudret(2011)]{Hyperbolic_Truncation_Scheme}
Géraud Blatman and Bruno Sudret.
\newblock Adaptive sparse polynomial chaos expansion based on least angle
  regression.
\newblock \emph{Journal of Computational Physics}, 230\penalty0 (6):\penalty0
  2345 -- 2367, 2011.
\newblock ISSN 0021-9991.
\newblock \doi{https://doi.org/10.1016/j.jcp.2010.12.021}.
\newblock URL
  \url{http://www.sciencedirect.com/science/article/pii/S0021999110006856}.

\bibitem[Yang et~al.(2016)Yang, Lei, Baker, and Lin]{PCE_Rotations}
Xiu Yang, Huan Lei, Nathan~A. Baker, and Guang Lin.
\newblock Enhancing sparsity of {H}ermite polynomial expansions by iterative
  rotations.
\newblock \emph{Journal of Computational Physics}, 307:\penalty0 94 -- 109,
  2016.
\newblock ISSN 0021-9991.
\newblock \doi{https://doi.org/10.1016/j.jcp.2015.11.038}.
\newblock URL
  \url{http://www.sciencedirect.com/science/article/pii/S0021999115007780}.

\bibitem[Pan and Dias(2017)]{PCE_Linear_Combination_of_Variables}
Qiujing Pan and Daniel Dias.
\newblock Sliced inverse regression-based sparse polynomial chaos expansions
  for reliability analysis in high dimensions.
\newblock \emph{Reliability Engineering \& System Safety}, 167:\penalty0 484 --
  493, 2017.
\newblock ISSN 0951-8320.
\newblock \doi{https://doi.org/10.1016/j.ress.2017.06.026}.
\newblock URL
  \url{http://www.sciencedirect.com/science/article/pii/S095183201630864X}.
\newblock Special Section: Applications of Probabilistic Graphical Models in
  Dependability, Diagnosis and Prognosis.

\bibitem[Hastie and Tibshirani(2010)]{Regularized_Linear_Regression}
Trevor Hastie and Rob Tibshirani.
\newblock Regularization paths for generalized linear models via coordinate
  descent.
\newblock \emph{Journal of Statistical Software}, 33:\penalty0 1 -- 22, 2010.
\newblock ISSN 1548-7660.
\newblock URL \url{https://www.ncbi.nlm.nih.gov/pubmed/20808728}.
\newblock Special Section: Applications of Probabilistic Graphical Models in
  Dependability, Diagnosis and Prognosis.

\bibitem[Guo et~al.(2019)Guo, Liu, and Zhou]{GUO2019129}
Ling Guo, Yongle Liu, and Tao Zhou.
\newblock Data-driven polynomial chaos expansions: A weighted least-square
  approximation.
\newblock \emph{Journal of Computational Physics}, 381:\penalty0 129 -- 145,
  2019.
\newblock ISSN 0021-9991.
\newblock \doi{https://doi.org/10.1016/j.jcp.2018.12.020}.
\newblock URL
  \url{http://www.sciencedirect.com/science/article/pii/S0021999119300014}.

\bibitem[Kaintura et~al.(2018)Kaintura, Dhaene, and Spina]{PC_Hermite}
Arun Kaintura, Tom Dhaene, and Domenico Spina.
\newblock Review of polynomial chaos-based methods for uncertainty
  quantification in modern integrated circuits.
\newblock \emph{ELECTRONICS}, 7\penalty0 (3):\penalty0 21, 2018.
\newblock ISSN 2079-9292.
\newblock URL \url{http://dx.doi.org/10.3390/electronics7030030}.

\bibitem[Abramowitz and Stegun(1964)]{Book_on_Orthogonal_Polynomials}
Milton Abramowitz and Irene~A. Stegun.
\newblock \emph{Handbook of Mathematical Functions with Formulas, Graphs, and
  Mathematical Tables}.
\newblock Dover, New York, ninth dover printing, tenth gpo printing edition,
  1964.

\bibitem[Cortés et~al.(2017)Cortés, Romero, Roselló, and
  Villanueva]{CORTES20171}
J.-C. Cortés, J.-V. Romero, M.-D. Roselló, and R.-J. Villanueva.
\newblock Improving adaptive generalized polynomial chaos method to solve
  nonlinear random differential equations by the random variable transformation
  technique.
\newblock \emph{Communications in Nonlinear Science and Numerical Simulation},
  50:\penalty0 1 -- 15, 2017.
\newblock ISSN 1007-5704.
\newblock \doi{https://doi.org/10.1016/j.cnsns.2017.02.011}.
\newblock URL
  \url{http://www.sciencedirect.com/science/article/pii/S1007570417300588}.

\bibitem[Shiryaev(1996)]{Probability_Textbook}
Albert~N. Shiryaev.
\newblock \emph{Probability}.
\newblock Springer-Verlag New York, 1996.
\newblock ISBN 978-1-4757-2539-1.
\newblock \doi{https://doi.org/10.1007/978-1-4757-2539-1}.

\bibitem[Mol et~al.(2009)Mol, Vito, and Rosasco]{Elastic_Net1}
Christine~De Mol, Ernesto~De Vito, and Lorenzo Rosasco.
\newblock Elastic-net regularization in learning theory.
\newblock \emph{Journal of Complexity}, 25\penalty0 (2):\penalty0 201 -- 230,
  2009.
\newblock ISSN 0885-064X.
\newblock \doi{https://doi.org/10.1016/j.jco.2009.01.002}.
\newblock URL
  \url{http://www.sciencedirect.com/science/article/pii/S0885064X0900003X}.

\bibitem[Amaratunga and Cabrera(2001)]{Quantile}
Dhammika Amaratunga and Javier Cabrera.
\newblock Analysis of data from viral {DNA} microchips.
\newblock \emph{Journal of the American Statistical Association}, 96\penalty0
  (456):\penalty0 1161--1170, 2001.
\newblock \doi{10.1198/016214501753381814}.
\newblock URL \url{https://doi.org/10.1198/016214501753381814}.

\bibitem[Rosenblatt(1952)]{rosenblatt1952}
Murray Rosenblatt.
\newblock Remarks on a multivariate transformation.
\newblock \emph{Ann. Math. Statist.}, 23\penalty0 (3):\penalty0 470--472, 09
  1952.
\newblock \doi{10.1214/aoms/1177729394}.
\newblock URL \url{https://doi.org/10.1214/aoms/1177729394}.

\bibitem[Torre et~al.(2018)Torre, Marelli, Embrechts, and
  Sudret]{Modified_Ishigami}
E.~Torre, S.~Marelli, P.~Embrechts, and B.~Sudret.
\newblock Data-driven polynomial chaos expansion for machine learning
  regression.
\newblock \emph{arXiv preprint arXiv:1808.03216}, 2018.
\newblock URL \url{https://arxiv.org/abs/1808.03216}.

\bibitem[Manceau and Rohmer(2016)]{UQ_CO2_2}
J.~Manceau and J.~Rohmer.
\newblock Post-injection trapping of mobile {CO2} in deep aquifers: Assessing
  the importance of model and parameter uncertainties.
\newblock \emph{Computational Geosciences}, 20:\penalty0 1251 -- 1267, 2016.
\newblock ISSN 1573-1499.
\newblock \doi{https://doi.org/10.1007/s10596-016-9588-x}.

\bibitem[Efron et~al.(2004)Efron, Hastie, Johnstone, and Tibshirani]{efron2004}
Bradley Efron, Trevor Hastie, Iain Johnstone, and Robert Tibshirani.
\newblock Least angle regression.
\newblock \emph{Ann. Statist.}, 32\penalty0 (2):\penalty0 407--499, 04 2004.
\newblock \doi{10.1214/009053604000000067}.
\newblock URL \url{https://doi.org/10.1214/009053604000000067}.

\bibitem[Rubinstein et~al.(2008)Rubinstein, Zibulevsky, and
  Elad]{singaravelu:06:eurosys}
Ron Rubinstein, Michael Zibulevsky, and Michael Elad.
\newblock Efficient implementation of the {K}-{SVD} algorithm using batch
  orthogonal matching pursuit.
\newblock \emph{CS Technion}, 40, 01 2008.

\bibitem[Ackley(1987)]{Ackley_Function}
David Ackley.
\newblock \emph{A Connectionist Machine for Genetic Hillclimbing}.
\newblock Springer US, 1987.
\newblock ISBN 978-1-4612-9192-3.
\newblock \doi{https://doi.org/10.1007/978-1-4613-1997-9}.

\bibitem[Ma and Zabaras(2011{\natexlab{b}})]{MA20117311}
Xiang Ma and Nicholas Zabaras.
\newblock Kernel principal component analysis for stochastic input model
  generation.
\newblock \emph{Journal of Computational Physics}, 230\penalty0 (19):\penalty0
  7311 -- 7331, 2011{\natexlab{b}}.
\newblock ISSN 0021-9991.
\newblock \doi{https://doi.org/10.1016/j.jcp.2011.05.037}.
\newblock URL
  \url{http://www.sciencedirect.com/science/article/pii/S0021999111003494}.

\bibitem[Ahmed(2019)]{AHMED2019901}
Tarek Ahmed.
\newblock Chapter 14 - {P}rinciples of {W}aterflooding.
\newblock In Tarek Ahmed, editor, \emph{Reservoir Engineering Handbook (Fifth
  Edition)}, pages 901 -- 1107. Gulf Professional Publishing, fifth edition
  edition, 2019.
\newblock ISBN 978-0-12-813649-2.
\newblock \doi{https://doi.org/10.1016/B978-0-12-813649-2.00014-1}.
\newblock URL
  \url{http://www.sciencedirect.com/science/article/pii/B9780128136492000141}.

\bibitem[Henderson and Pena(2017)]{HENDERSON2017178}
Nélio Henderson and Luciana Pena.
\newblock Simulating effects of the permeability anisotropy on the formation of
  viscous fingers during waterflood operations.
\newblock \emph{Journal of Petroleum Science and Engineering}, 153:\penalty0
  178 -- 186, 2017.
\newblock ISSN 0920-4105.
\newblock \doi{https://doi.org/10.1016/j.petrol.2017.03.047}.
\newblock URL
  \url{http://www.sciencedirect.com/science/article/pii/S0920410517304102}.

\bibitem[Brooks and A.T.Corey(1964)]{Relative_Phase_Permeability}
R.H. Brooks and A.T.Corey.
\newblock Hydraulic properties of porous media.
\newblock \emph{Hydrology Papers}, \penalty0 (3), 1964.

\end{thebibliography}

\end{document}